\documentclass[manuscript]{emulateapj}

\slugcomment{to appear in the Astrophysical Journal}

\shorttitle{PU Vul--evolution of WD, RG, and nebula}
\shortauthors{Kato et al.}

\begin{document}


\title{Evolution of the Symbiotic Nova PU Vul -- Outbursting White Dwarf, 
 Nebulae, and Pulsating Red Giant Companion}


\author{Mariko Kato}
\affil{Department of Astronomy, Keio University, Hiyoshi, Yokohama
  223-8521 Japan}
\email{mariko@educ.cc.keio.ac.jp}

\author{Joanna Miko\l ajewska}
\affil{N. Copernicus Astronomical Center, Bartycka 18,00-716 Warszawa, Poland  }

\and 

\author{Izumi Hachisu}
\affil{Department of Earth Science and Astronomy, College of Arts and
Sciences, University of Tokyo, Komaba, Meguro-ku, Tokyo 153-8902, Japan}



\begin{abstract}
We present a composite light-curve model of the symbiotic nova PU Vul 
(Nova Vulpeculae 1979) 
that shows a long-lasted flat optical peak followed by a slow decline. 
Our model light-curve consists of
three components of emission, i.e., an outbursting white dwarf (WD), 
its M-giant companion, and nebulae. 
The WD component dominates in the flat peak while the nebulae dominate  
after the photospheric temperature of the WD rises to 
$\log T$ (K) $\gtrsim 4.5$, suggesting its WD origin. 
We analyze the 1980 and 1994 eclipses to be 
total eclipses of the WD occulted by the pulsating M-giant companion with  
two sources of the nebular emission; 
one is an unocculted nebula of the M-giant's 
cool-wind origin and the other is a partially occulted nebula associated 
to the WD. 
We confirmed our theoretical outburst model of PU Vul by 
new observational estimates, that spanned 32 yr, of the temperature 
and radius. Also our eclipse analysis confirmed that the WD photosphere 
decreased by two orders of magnitude between the 1980 and 1994 eclipses.
We obtain the reddening $E(B-V) \sim 0.3$ and distance to PU Vul $d \sim 4.7$ kpc. 
We interpret the recent recovery of brightness in terms of eclipse of 
the hot nebula surrounding the WD, suggesting that hydrogen burning 
is still going on.  
To detect supersoft X-rays, we recommend X-ray observations around 
June 2014 when absorption by neutral hydrogen is minimum. 
\end{abstract}




\section{Introduction} \label{sec_introduction}

Symbiotic novae are thermonuclear runaway phenomena occurring 
on white dwarfs (WDs) in 
binary systems that consist of a WD and a red giant (RG). 
Symbiotic novae can be divided into two subgroups according to 
their spectral evolutions. The first group exhibit a long 
(several years) ``supergiant phase,'' resembling an A-F supergiant  
when the star underwent an outburst.   
In the second group, a nebular phase begins almost immediately 
after the optical maximum, and a ``supergiant phase,'' if there is, 
has a very short duration \citep{mue94}. 
The first subgroup include AG Peg, RT Ser, RR Tel, and PU Vul. 
The second subgroup include V1016 Cyg, HBV 475, and HM Sge. 
It is, however, still unknown the reason why this difference arises. 
Due to very long evolution-timescales (one to several tens of years or more) 
it is not easy to obtain observational data of a whole period of the outbursts, 
such as dense and continuous photometric,  
spectroscopic, and multi-wavelength observations including UV and X-ray.    
Under these circumstances, 
it has been difficult to study symbiotic novae 
quantitatively compared with classical novae.
 
Among symbiotic novae, PU Vul is a rare exception. 
It is an eclipsing binary of the orbital period $\sim 4900$ days (13.4 yr)     
 \citep{kol95,nus96,gar96,shu11}.  
During eclipses, different emission components are occulted 
differently. This offers
 a good chance for quantitative study. 
PU Vul outbursted in 1979 and we have dense optical spectroscopic/photometric 
data as well as {\it IUE/HST} UV observations.
Recently, \citet{kat11} first presented a 
theoretical model of PU Vul that reproduces the optical flat peak 
as well as the UV light curve, and estimated   
the WD mass ($\sim 0.6~M_\odot$). Their model is, however, only 
for the light curve of the outbursting component (WD) 
and the other emission components were neglected. 

This paper presents a comprehensive model of emission components 
of the WD, RG, and nebulae, 
based on new estimates of the temperature and radius of the hot component (WD),
as well as the cool component (RG) derived from the two eclipses 
(1980 and 1994).  
Section \ref{sec_model} briefly introduces our evolution model of PU Vul. 
Using our theoretical light curves, we constrain the  
extinction and distance to PU Vul in Section \ref{sec_EandD}. 
Section \ref{section_TR} compares our theoretical light curves 
with our new observational estimates of temperature and radius of the 
WD.  
In Section \ref{sec_eclipse}, we analyze light curves of the two eclipses 
and obtain the binary parameters as well as the brightnesses of the 
RG and nebulae. Using these values, we construct 
a composite light curve model of PU Vul in 
Section \ref{sec_opticallightcuve}. 
Discussion and conclusions follow in Sections \ref{sec_discussion} and 
\ref{sec_conclusions}.

\section{Model of WD component} \label{sec_model}

\subsection{Evolution of Nova Outbursts} \label{subsec_model}

A nova is a thermonuclear runaway event on a WD. After the hydrogen 
shell flash sets in, the envelope on the WD expands to a giant size. 
After it reaches the optical peak, the envelope settles 
down into a steady-state. 
The optical magnitude decays as the envelope mass decreases while 
the photospheric temperature rises with time. 
The decay phase can be followed by a
quasi-static sequence \citep{kat11}. 
We solved the equations of hydrostatic balance, continuity, 
radiative diffusion, and 
conservation of energy, from the bottom of the hydrogen-rich 
envelope through the photosphere.  
The evolution is followed by a sequence of decreasing envelope mass. 
The time interval $ \Delta t$ between two successive solutions  
is calculated by
 $\Delta t= \Delta M_{\rm env} / (\dot M_{\rm nuc} + \dot M_{\rm wind})$, 
where $\Delta M_{\rm env}$ is the difference between the envelope masses 
of the two successive solutions, and $\dot M_{\rm nuc}$ is the 
hydrogen nuclear burning rate  and $\dot M_{\rm wind}$ is  
the optically-thin wind mass-loss rate  
\citep[see Equation (24) in][for more detail]{kat94h}.
The method and numerical techniques are essentially the same as those  
in \citet{kat11}. We used OPAL opacities \citep{igl96}.
The WD radius (the bottom of hydrogen shell-burning) 
is assumed to be the Chandrasekhar radius. 
The mixing-length parameter of convection $\alpha$ is assumed to be 1.5 
(see \citet{kat09} for the dependence of the light curve on $\alpha$).  
Internal structures of the envelope are essentially the same as those in  
Figure 7 of \citet{kat11}. 
We calculate optical and UV light curves from the blackbody spectrum  
with the photospheric temperature, $T_{\rm ph}$. 
To calculate $V$ magnitude, we use 
the standard Johnson $V$ bandpass and add a bolometric correction of 
0.17 mag (see Section \ref{section_TR}). 

In our model, we simply assume uniform chemical composition of the envelope. 
PU Vul does not show any CO/Ne enrichment  
but the overall chemical composition is 
almost consistent with being solar; slightly subsolar of iron 
\citep{bel84,bel89} and helium overabundance \citep{and94,lun05} are reported. 
Thus, we assume four different sets of chemical composition ($X$, $Y$, $Z$)
by weight for hydrogen, helium, and heavy elements of the envelope as   
(0.7, 0.29, 0.01), (0.7, 0.28, 0.02) (0.5, 0.49, 0.01), and (0.5, 0.49, 0.006). 
Here $Z=0.01$ is closer to the recent estimate of heavy element abundance of 
solar composition \citep[$Z=0.0128$:][]{gre08}. 
The WD mass is assumed to be 0.6 $M_\odot$ as listed in Table \ref{table_model}. 
Model 4 in Table \ref{table_model} is the same as Model 2 in \citet{kat11}. 

A typical classical nova shows heavy element enrichment (C, O, and Ne) 
in its ejecta, which is interpreted in terms of dredge-up of WD material
\citep{pri84,pri95}. 
PU Vul shows no indication of such enhancement in spectra, 
which suggests that the WD is not eroded during and before the outburst. 
The theoretical model described in \citet{kat11} showed that only a small part of the 
accreted matter was lost in the optically-thin wind, 
and the rest was burned to helium due to hydrogen nuclear burning 
and accumulated on the WD.   
Therefore, the WD develops a helium layer underneath the newly 
accreted material. 
In the next outburst, a part of the helium layer will possibly be dredged up 
and mixed into the upper hydrogen layer. In such a case the envelope 
will become helium-rich like in Models 3 and 4 in Table \ref{table_model}.

There are observational evidences of wind mass-loss from WDs  
in some symbiotic stars. 
For PU Vul, \citet{tom91} found broad emission wings in \ion{H}{1}, \ion{He}{1},
\ion{He}{2} and \ion{N}{4} lines as well as violet-shifted P Cygni type 
absorption components in \ion{H}{1} and \ion{He}{1} lines in the 
optical spectra taken in 1990-91, which they attributed to the hot component 
winds. \citet{sio93} discussed the onset of Wolf-Rayet type wind outflowing 
from the hot component based on the {\it IUE} high resolution spectra of 
1989-1991, and estimated an upper limit of 
$\dot M_{\rm wind} \lesssim 10^{-5}~M_\odot$~yr$^{-1}$. 
For AG Peg, the outburst lasted about 150 yr, which suggests 
a low mass WD with no optically-thick winds. The wind mass-loss 
rate from the hot component was estimated to be of the order of 
$10^{-7}~M_\odot$~yr$^{-1}$ \citep{vog94} and 
$10^{-6}~M_\odot$~yr$^{-1}$ \citep{ken93}. 
The intensity of the wind diminished in step with 
the hot component luminosity during the decline of the outburst. 
For AE Ara, the wind mass loss rate is estimated to be a few times $10^{-8}$ 
-- $ 10^{-7} M_\odot$~yr$^{-1}$ and the WD mass to be $M_{\rm h} \sin~i
\sim 0.4~M_\odot$ \citep{mik03}.  

With such poor information on mass-loss rates, we simply assume 
that an optically-thin wind begins to blow when the photospheric temperature 
rises to $\log T_{\rm ph}$ (K) $\sim 4.0$ and the wind continues until 
$\log T_{\rm ph}$ (K) $ \sim 5.05$ at various rates listed in 
Table \ref{table_model}  
(e.g., $\dot M_{\rm wind} = 5.0 \times 10^{-7} M_\odot$ yr$^{-1}$ for Model 1).  
After the temperature reaches 
$\log T_{\rm ph}$ (K) $ \sim 5.05$, the wind mass-loss rate drops to 
$\dot M_{\rm wind} \sim 1.0 \times 10^{-7} M_\odot$ yr$^{-1}$.

We cannot accurately determine the WD mass of PU Vul 
only from our light curve analysis.  \citet{kat11} obtained a plausible 
range of the WD mass, 0.5 -- 0.72 $M_\odot$ 
corresponding to a reasonable range of the wind mass-loss rates. 
In the present paper, we adopt an 
0.6 $M_\odot$ WD as a standard model of PU Vul 
(see Section \ref{subsection_UV} for more detail).

\placetable{table_model}
\begin{deluxetable*}{llccccl}
\tabletypesize{\scriptsize}
\tablecaption{Model Parameters  
\label{table_model}}
\tablewidth{0pt}
\tablehead{
\colhead{Subject} &
\colhead{} &
\colhead{Model 1} &
\colhead{Model 2} &
\colhead{Model 3} &
\colhead{Model 4} &
\colhead{Units}
}
\startdata
$X$ & ...    &0.7  &0.7  &0.5  & 0.5  \\
$Y$ & ...    &0.29 &0.28 &0.49 &0.494 \\
$Z$ & ...    &0.01 &0.02 &0.01 &0.006&\\
WD mass & ...&0.6  & 0.6 &0.6  &0.6& $M_\odot$ \\
$M_{\rm bol}$\tablenotemark{a}& ...&$-5.44$  &$-5.36$  &$-5.63$& $-5.68$& mag \\
$M_{V, {\rm peak}}$\tablenotemark{b}& ...& $-5.61$  & $-5.53$& $-5.80$ &$-5.85$& mag \\
$L_{\rm peak}$\tablenotemark{a} &...     &4.6 &4.2 &5.5&5.7  &  $10^{37}$erg~s$^{-1}$ \\
maximum radius \tablenotemark{c}& ... & 63    &60 &61 &64   &  $R_\odot$ \\
initial envelope mass & ...           & 4.0 &2.6 & 3.4 &4.6   & $10^{-5}~M_\odot$ \\
H-burning rate\tablenotemark{d}& ...  & 1.7 &1.6& 2.9 &3.0  &  $10^{-7}~M_\odot$~yr$^{-1}$ \\
assumed wind mass-loss rate ($T<5.05$)\tablenotemark{e}& ...&5.0 &3.0&2.0&3.0  & $10^{-7}~M_\odot$~yr$^{-1}$  \\
assumed wind mass-loss rate ($T>5.05$)\tablenotemark{f}& ...& 1.0&1.0&1.0&1.0 & $10^{-7}~M_\odot$~yr$^{-1}$  
\enddata
\tablenotetext{a}{Typical values of the optical flat peak at $\log T_{\rm ph}$ (K) =3.9.}
\tablenotetext{b}{We adopt $M_{\rm bol}-0.17$ mag.}
\tablenotetext{c}{The radius reached before $\log T_{\rm ph}$ (K) =3.9.}
\tablenotetext{d}{Values at $\log T_{\rm ph}$ (K) =4.5.}
\tablenotetext{e}{Optically-thin wind from $\log T_{\rm ph} $(K) = 4 to 5.05.} 
\tablenotetext{f}{Optically-thin wind from $\log T_{\rm ph} $(K) = 5.05 to 
                 the end of hydrogen burning.}
\end{deluxetable*}

\begin{figure}
\epsscale{1.15}
\plotone{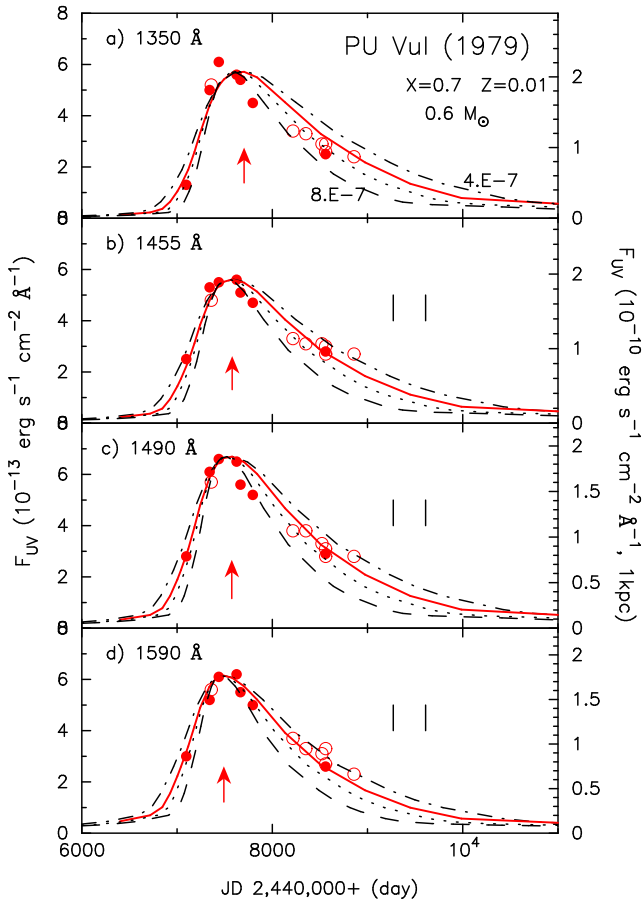}
\caption{ UV light curves for four narrow bands at,  
a) 1350~\AA, b) 1455~\AA, c) 1490~\AA, and d) 1590~\AA.
Theoretical light curves are also shown for an assumed distance of 1 kpc 
(with right-side axis) and no absorption; They are 0.6 $M_\odot$ WDs with the 
chemical composition of the envelope $X=0.7$ and $Z=0.01$ 
with four different optically-thin wind 
mass-loss rates;     
Dash-dotted curves: $4\times 10 ^{-7}M_\odot$yr$^{-1}$.   
Solid curves: $5\times 10 ^{-7}M_\odot$yr$^{-1}$ (Model 1).   
Dotted curves: $6\times 10 ^{-7}M_\odot$yr$^{-1}$. 
Dashed curves: $8\times 10 ^{-7}M_\odot$yr$^{-1}$.
Short vertical lines in panels b), c), and d) show the period of the second 
eclipse at the optical $V$ band between JD 2,449,270 and 2,449,610.   
Among the observational data, the 
open circles denote the ones with low accuracy because they were observed with 
a short exposure time ($< 1000$ s) or it is obtained from very noisy spectra. 
The red arrows indicate the epoch of the UV maximum in each wavelength band.  
}
\label{lightUV4.f}
\end{figure}

\subsection{Continuum UV Light Curve} \label{subsection_UV}

In classical novae, a narrow spectral region around 1455~\AA~ is known to be 
emission-line free and can be a representative 
of continuum flux \citep{cas02}. This continuum band has been used to determine 
distances to several classical novae \citep{hac06,hac08,kat09v838her}, 
and also used in analysis of PU Vul \citep{kat11}. 
As PU Vul shows much weaker emission lines in its spectra than 
classical novae, 
we can use three other wavelength bands around 
1350, 1490 and 1590~\AA~ of a 20 ~\AA~ width, in addition to the 
 UV 1455 ~\AA~ band.
Figure \ref{lightUV4.f} depicts light curves of these four narrow bands, 
extracted from the {\it IUE} data archive\footnote{http://sdc.laeff.inta.es/ines/}. 

During the outburst, the photospheric temperature gradually rises and the
photospheric radius shrinks while the bolometric luminosity is almost constant. 
Thus, a UV light curve has the peak at a certain temperature. 
Figure \ref{lightUV4.f} also shows theoretical UV light curves that represent 
continuum emission in each wavelength. 
These four light curves show basically a similar behavior, because each 
wavelength is close. In a shorter wavelength band, the UV flux reaches maximum  
slightly later than in the other longer bands as indicated by upward arrows.  

The flux at the observed peak is obtained to be 
 $F_{1350}=5.7\times 10^{-13}$, $F_{1455}=5.6 \times 10^{-13}$, 
$F_{1490}=6.7 \times 10^{-13}$, and  
$F_{1590}=6.1\times 10^{-13}$ erg~s$^{-1}$cm$^{-2}$\AA$^{-1}$, respectively.
If the emission can be approximated by blackbody, unabsorbed peak fluxes should 
be larger in a shorter wavelength band, while the absorbed fluxes are in 
the inverse order. Comparing these peak fluxes, we see that 
the 1455~\AA~ band flux is too small, because the peak flux is more 
absorbed by cool winds from the M giant companion 
than in the other three bands of 1350, 1490 and 1590 ~\AA~ \citep{sho93}. 
The excess of $F_{1490}$ may be explained by contamination of emission lines. 
Considering these effects, we use the 1590 ~\AA~ band  
in the following discussion.

Figure \ref{lightUV4.f} also shows model light curves of the 0.6 $M_\odot$ WD 
with the chemical 
composition of $X=0.7,~Y=0.29$, and $Z=0.01$.  Each band light curve is  
made from blackbody emission of our evolution model. 
Here, we assume four optically-thin 
wind mass-loss rates of 4--8 $\times 10^{-7}~M_\odot$~yr$^{-1}$.
For a higher mass-loss rate, the evolution is faster and the UV light curve shape is 
narrower. All these light curves more or less agree with the observational 
UV light curve in each wavelength band, and we chose the 
$5\times 10^{-7}~M_\odot$~yr$^{-1}$
as having the best agreement with these data points.  

For a given chemical composition \citet{kat11} obtained a range 
of the WD mass that reasonably well reproduces the UV light 
curve for reasonable rates of the optically-thin mass-loss. 
The lowest WD mass is obtained for a very large wind mass-loss 
rate of $1 \times 10^{-6}M_\odot$ yr$^{-1}$, while the highest WD 
mass is for no wind mass-loss. 
For example, if we fix the chemical composition to be $X=0.7$ and 
$Z=0.01$, a plausible WD mass is between 0.52 and 0.72 $M_\odot$, 
corresponding to the wind mass-loss rate of $1 \times 10^{-6}M_\odot$ yr$^{-1}$ 
and no mass-loss, respectively.  
These ranges of the WD mass are summarized in Table \ref{table_distancemodulus}
for four specified chemical compositions.  
This table also shows a range of the bolometric luminosities at the optical flat peak. 
The larger the bolometric luminosity, the more massive the WD.
Combining these theoretical bolometric luminosities with the observed magnitudes, 
we can derive a range of the distance moduli, $(m-M)_V$, which are shown in the last 
column of Table \ref{table_distancemodulus}.

\placetable{table_distancemodulus}
\begin{deluxetable*}{llccllll}
\tabletypesize{\scriptsize}
\tablecaption{Range of Distance Moduli
\label{table_distancemodulus}}
\tablewidth{0pt}
\tablehead{
\colhead{Composition} &
\colhead{} &
\colhead{WD Mass\tablenotemark{a}} &
\colhead{$L_{\rm bol}$\tablenotemark{b}} &
\colhead{$V_{\rm peak}-M_{\rm V}$\tablenotemark{c}} &
\colhead{}\\
\colhead{$(X,Y,Z)$} &
\colhead{} &
\colhead{($M_\odot$)}&
\colhead{($10^{37}$erg~s$^{-1}$)} &
\colhead{}
}
\startdata
(0.7,~0.29,~~0.01)&...& 0.52 -- 0.72 & 3.4 -- 6.1 & 13.87 -- 14.52\\
(0.7,~0.28,~~0.02)&...& 0.5 -- 0.67 & 3.0 -- 5.0 & 13.75 -- 14.30\\
(0.5,~0.49,~~0.01)&...& 0.5  -- 0.62  & 3.0 -- 5.7& 14.03 -- 14.45\\
(0.5,~0.494,~0.006)&...& 0.53  -- 0.65 & 4.5 -- 6.5& 14.19 -- 14.58\\
\enddata
\tablenotetext{a}{A range of the WD mass obtained from the UV light curve fitting. 
The lower limit corresponds to the case of a very large mass-loss rate of 
$1\times 10^{-6}~M_\odot~$yr$^{-1}$, while the upper limit is the
extreme case of no wind mass-loss \citep[see][]{kat11}.}
\tablenotetext{b}{Values at $\log T$ (K)=3.90.}
\tablenotetext{c}{We adopt $V_{\rm peak}=8.59$ for the optical flat peak.
Theoretical absolute $V$-magnitudes are calculated as $M _{\rm V}=M _{\rm bol}-0.17$.}
\end{deluxetable*}

\section{Extinction and Distance}  \label{sec_EandD}

Before deriving physical parameters of the nova, we must 
estimate the extinction and distance. 
The reddening was estimated by various methods, 
\ion{H}{1} Balmer line ratios, \ion{He}{2} emission line ratios, 
interstellar optical/UV absorption features, and comparison between the observed 
optical/near-IR spectra and some standards 
\citep{bel82b,bel84,fri84,ken86,goc91,vog92,hoa96,rud99,lun05}. 
They are unfortunately scattered in a broad range of $E(B-V)=0.22-0.53$.  
Thus, we have 
made our own estimates based on the theoretical light curves (Section \ref{subsec_E(B-V)}) 
and comparison between spectral classification and colors 
(Section \ref{subsec_color}). 

\begin{figure*}
\epsscale{0.8}
\plotone{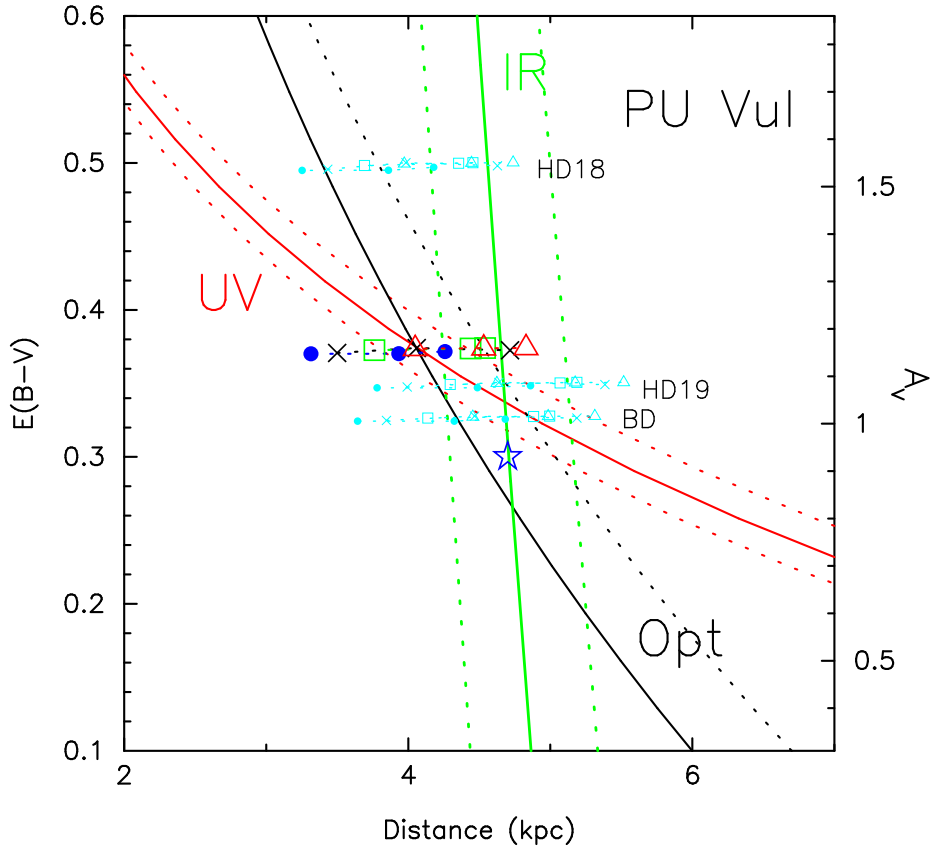}
\caption{
Distance-reddening relation of PU Vul. 
The distance-reddening relations obtained from the    
optical light curve fitting of Model 1; black solid line: $(m-M)_V=14.20$,  
i.e., Equation (\ref{equation_distmod_optical2}); black dotted line: $(m-M)_V=14.44$.
The red solid line represents the UV 1590~ \AA~  light curve 
fitting of Model 1 (Equation (\ref{equation_distmod_UV1590_2})).
The two red dotted lines beside the line represent possible $\pm 15 \%$ errors in 
absolute UV flux.  
The distance-reddening relation derived from the K-magnitude fitting 
(Equation (\ref{equation_distmod_IR2})) 
is depicted by the green solid line with $\pm 0.2$ mag error lines in both sides. 
The central black cross indicates the intersection of the UV and optical fluxes 
of Model 1. 
The crosses in both sides represent the intersections for $M_{\rm WD}=0.52$ (left) 
and 0.72 $M_\odot$ (right) WD models with the chemical composition of  
$X=0.7$ and $Z=0.01$, 
i.e., the lowest and highest WD masses in Table \ref{table_distancemodulus}. 
Other models are depicted by the different symbols: blue filled circles for 
Model 2, green open squares for Model 3, red triangles for Model 4.   
The intersections for the same models but for different extinction law are 
also plotted by blue symbols: ($R_V,~ R_{1590}$)=(2.40, 6.29) for HD185418, 
(2.48, 8.03) for HD197702, and (2.91, 8.84) for BD +35 4258. 
The same symbol indicates the same chemical composition and the middle point  
of the three same symbols corresponds to the models in Table \ref{table_model}. 
The name of each star is indicated beside the corresponding group of points 
with first few letters.   
The open star mark represent $d$=4.7 kpc and $E(B-V)=0.3$. 
See Section \ref{subsec_E(B-V)} for more detail. 
\label{pu_vul_extinction}
}
\end{figure*}

\placefigure{pu_vul_extinction}

\subsection{Extinction from Model Light Curves}  \label{subsec_E(B-V)}

From our light curve fittings, we get relations on the extinction $E(B-V)$ 
and the distance $d$ to PU Vul. 
The distance modulus is 
\begin{equation}
(m-M)_V= A_V  + 5 \log ~(d/1~{\rm kpc}) +10,  
\label{equation_distmod_optical}
\end{equation}
where $A_V=R_V \times E(B-V)$ and $R_V=3.1$. 
In the optical maximum phase, 1979-1986, except the eclipse, the mean 
magnitude is obtained to be $V=8.59 \pm 0.06$ (see Table \ref{table_TandRofWD}), 
whereas  
the absolute bolometric magnitude is $M_{\rm bol}=-5.44$ from Model 1 
(Table \ref{table_model}).
Here, we adopt a bolometric correction of BC($V$)=0.17 (see Section \ref{section_TR}),
 as a representative value for an extended WD photosphere during the 
A-F spectral phase. Then, we have
\begin{equation}
14.20 = 3.1 \times E(B-V) + 5 \log ~(d/1~{\rm kpc}) + 10.
\label{equation_distmod_optical2}
\end{equation}
This equation gives a relation between $E(B-V)$ and $d$ for a specified model, Model 1, 
which is depicted in Figure \ref{pu_vul_extinction}.    
There is another possible way of fitting. In 1979, PU Vul showed a spectral type of 
F0 I without emission lines, and its magnitude was about $V=8.87$ 
(see Figure \ref{light}).
If we take a bolometric correction typical for F0 I/II, BC($V$)=0.13 \citep{str81},
we have a larger distance modulus $(m-M)_V=14.44$ for the same Model 1. 
This case is also plotted in Figure \ref{pu_vul_extinction}.

We have another distance-reddening relation from the UV 1590 ~\AA\ light curve 
fitting, i.e., 
\begin{eqnarray}
-2.5~ \log F_{1590}({\rm obs}) & = & -2.5~\log F_{1590}({\rm model}) \cr
& & +A_\lambda + 5~\log ~(d/{\rm 1~kpc}).
\label{equation_distmod_UV1590}
\end{eqnarray}
Here $A_\lambda = R_{1590}~E(B-V)$ and we adopt $R_{1590}$=8.3 \citep{fit07}. 
\citet{sea79}'s formula gives a similar value of $7.9$. 
Figure \ref{lightUV4.f}(d) shows 
$F_{1590}({\rm obs})=6.1 \times 10^{-13}$ erg~s$^{-1}$~cm~$^{-2}~$~\AA$^{-1}$ at the 
UV 1590 ~\AA~ peak,  
whereas  
$F_{1590}({\rm model})=1.77 \times 10^{-10}$ erg~s$^{-1}$~cm~$^{-2}~$\AA$^{-1}$
with an assumed distance of $d=1$ kpc. 
Substituting these values into equation (\ref{equation_distmod_UV1590}), we get 
a relation 
\begin{equation}
  6.15 = 8.3 \times E(B-V) + 5 \log ~(d /1~{\rm kpc})  
\label{equation_distmod_UV1590_2}
\end{equation}
for Model 1. Figure \ref{pu_vul_extinction} also shows Equation 
(\ref{equation_distmod_UV1590_2}) with two additional lines in the both sides which 
represent a possible $15 \%$ error in the light curve fitting. This error is 
a summation of the accuracy of the absolute flux calibration of {\it IUE} ($\sim 5 \%$)
and possible contamination of emission/absorption line contribution in the region of  
1590 ~\AA, which we assumed to be 10 \%.   

Combined these two fittings, i.e., Equations (\ref{equation_distmod_optical2})
and (\ref{equation_distmod_UV1590_2}), we obtain $E(B-V)=0.37$ and $d=4.1$ kpc,  
which are plotted by a black X-mark (the middle one among the three Xs). 
If we assume a different WD mass, we get a different relation between  
$E(B-V)$ and $d$, because $M_V$ is different for a different WD mass model. 
The two X-marks in the left/right sides in Figure \ref{pu_vul_extinction} indicate 
the intersection of the two extreme cases 
of $M_{\rm WD}=0.52$ and 0.72 $M_\odot$, corresponding  
to the lowest and highest WD masses (see Table \ref{table_distancemodulus}). 

For different sets of chemical composition, we also get different intersections  
which are shown by different symbols in Figure \ref{pu_vul_extinction}.  
From these points we see that $E(B-V)=0.37$ is almost independent of 
the WD mass or chemical composition. 
This is because we use the same response (passband) functions to derive 
$M_V$ and $\log F_{1590} ({\rm model})$ from blackbody spectrum of each model, and 
therefore, the ratios of the two values are common in all the models.  
As a result, these two equations yield a common value of $E(B-V)$ 
independent of the model. On the other hand, the distance depends on the WD mass 
and chemical composition ($X$), because a more massive WD/smaller $X$ has a larger  
photospheric luminosity, which results in a larger distance.  
In this way, we could not determine the distance only from the light curve fittings. 
We can constrain the distance corresponding to a permitted range of the WD mass  
as listed in Table \ref{table_model}.

It should be noted that Galactic interstellar absorption has very large 
uncertainty around the average value we adopted \citep[see e.g.][]{fit07}. 
Unfortunately, the extinction curve are not known in PU Vul, nor 
in the stars close to its same sight line. 
The closest stars are relatively far away; towards HD 185418 and HD 197702 , 
both $\sim 11 \deg$ away from PU Vul, 
and towards BD +35 4258, $\sim 15 \deg$ away. The values of   
($R_V$ and $R_{1590}$) are (2.40, 6.29) for HD 185418, 
(2.48, 8.03) for HD 197702, and (2.91, 8.84) for BD +35 4258. 
Using these values, we obtain the intersection from  
 Equations (\ref{equation_distmod_optical}) and (\ref{equation_distmod_UV1590}), 
which are also shown in Figure \ref{pu_vul_extinction}. 
These extinction estimates strongly depend on the adopted extinction 
curve. 

Thus, we could not accurately determine the distance 
and $E(B-V)$ from the light curve fittings of UV/optical bands.    

\placetable{table_reddening}
\begin{deluxetable*}{lcccccccc}
\tabletypesize{\scriptsize}
\tablecaption{Color excess in optical 
\label{table_reddening}}
\tablewidth{0pt}
\tablehead{
\colhead{Date} &
\colhead{Sp. Type\tablenotemark{a}} &
\colhead{$<V>$} &
\colhead{$<B-V>$\tablenotemark{b}} &
\colhead{$(B-V)_0$\tablenotemark{c}} &
\colhead{$E(B-V)$} & 
\colhead{$<U-B>$\tablenotemark{b}} &
\colhead{$(U-B)_0$\tablenotemark{c}} &
\colhead{$E(U-B)$}
}
\startdata
Apr 1979 & A7\,II & 8.84 & $0.39 \pm 0.01$ & 0.12 & 0.27 & $0.24\pm0.01$ & 0.02 & 0.25\\
May 1979 & F0\,I & 8.93 & $0.42 \pm 0.01$ & 0.19 & 0.23 & $0.29\pm0.01$ & 0.16 & 0.13\\
May 1979\tablenotemark{d} &  F0\,I & 8.98 & $0.51\pm0.02$ & 0.19 & 0.32 & $0.43\pm0.03$ & 0.16 & 0.27\\
Jul 1979 & F0\,I & 8.84 & $0.44\pm0.01$ & 0.19 & 0.25 & $0.34\pm0.01$ & 0.16 & 0.18 \\
Sep 1979 & F5\,I & 8.78 & $0.57\pm0.01$ & 0.40 & 0.17 & $0.37\pm0.02$ & 0.32 & 0.05\\
Dec 1979 & F3\,I  & 8.78 & $0.69\pm0.01$ & 0.31 & 0.38 & 0.50 & 0.27 & 0.23\\  
Aug 1981 & F5\,I & 8.55 & $0.75 \pm 0.01$ & 0.40 & 0.35 & $0.47\pm0.01$ & 0.32 & 0.15\\
Aug 1981\tablenotemark{d} & F5\,I & 8.64 & $0.71\pm0.01$ & 0.40 & 0.31 & $0.55\pm0.01$ & 0.32 & 0.23\\
Jun 1982 & F5\,I & 8.50 & $0.78\pm0.01$ & 0.40 & 0.38 & $0.44\pm 0.02$ & 0.32 & 0.12\\
Sep 1982  & F5--8\,I & 8.67 & $0.74 \pm 0.01$ & 0.49 & 0.25 & $0.42 \pm 0.02$ & 0.36 & 0.06\\
Nov 1982 & F3--4\,I & 8.40 & $0.57\pm 0.01$ & 0.33 & 0.24 & $0.19\pm0.01$ & 0.28 & -0.09\\
Dec 1982 & F0\,I & 8.43 & $0.58\pm 0.01$ & 0.19 & 0.39 & $0.23\pm0.03$ & 0.16 & 0.07\\
Jul 1983 & F0\,I & 8.39 & $0.61\pm0.01$ & 0.19 & 0.42 & $0.27\pm0.01$ & 0.16 & 0.11\\
\enddata
\tablenotetext{a}{These spectral types are taken from \citet{kol83}, \citet{bel84}, and 
  \citet[and references therein]{mue94}.} 
\tablenotetext{b}{The average $B-V$ and $U-B$ color indices are calculated using 
  the data from \citet{bel82a,bel85,bel90,kol83,marg79,mahra79,whit79,bruch80}.}
\tablenotetext{c}{The intrinsic color indexes of supergiants, $(B-V)_0$ and 
     $(U-B)_0$, are from \citet{str92}.}
\tablenotetext{d}{The $UBV$ colors in May 1979 \citep{cho81} and in August 
    1981 \citep{pur82} show some systematic offset with respect 
    to those from  other sources, and are displayed separately.}
\end{deluxetable*}

\subsection{Optical Color Excess}\label{subsec_color}

A direct estimate of the color excess $E(B-V)$ can be derived from the 
comparison of observed $B-V$ colors and spectral types of the hot component 
during 1979-1983, i.e., at the optical maximum when strong emission lines were still absent. 
In particular, we have calculated one-month averaged $B-V$ and $U-B$ colors from 
the published photometry 
\citep{bel82a,bel85,bel90,kol83,marg79,mahra79,whit79,bruch80,cho81, pur82}  
for several epochs when the spectral classification of 
PU Vul was made \citep[and references therein]{kol83,bel84,mue94}. 
Table \ref{table_reddening} shows the one-month averaged $B-V$ and $U-B$ colors and 
spectral classification corresponding to each epoch.
Assuming that the nova envelope of PU Vul had a typical spectral type, 
we can estimate the extinction with the comparison to  
the intrinsic color index  corresponding to its spectral type \citep{str92}. 
The $B-V$ and $U-B$ colors of PU Vul are in good agreement with those of 
bright supergiants for 1979--81, 
and the average ratio of $E(U-B)/E(B-V)=0.66^{+0.12}_{-0.14}$ estimated 
for this period agrees with the predicted value of $\sim 0.63-0.69$ \citep{cmm89}.
Since 1982 there is some contamination by the nebular continuum emission, 
especially in the $U-B$ color. We thus obtain $E(B-V)=0.30 \pm 0.02$ for 
the whole 1979-1983 period, and $E(B-V)=0.29 \pm 0.02$ for 1979-81.

This value is in good agreement with the color excesses estimated from 
various emission line ratios. 
We estimate $E(B-V)\sim 0.2$--0.3, from our measurements 
of the \ion{He}{2}\,1640 emission line fluxes from {\it IUE} spectra obtained 
in October 1991 and August 1992 combined with the optical \ion{He}{2}\,4686 
line fluxes for the same epochs \citep{and95}, assuming electron temperature 
between 10\,000--20\,000\,K.
Similarly, \citet{rud99} derived  $E(B-V)=0.22 \pm 0.10$ from the \ion{O}{1} 
line ratios whereas \citet{lun05} reported $E(B-V)=0.29$ resulting from the 
Balmer \ion{H}{1} line ratios.

Our extinction estimates agree with estimates for the total Galactic 
extinction towards PU Vul. We estimated  $E(B-V)=0.35$ from the 
Galactic extinction distribution based on {\it COBE} and {\it IRAS} maps combined with the 
Leiden-Dwingeloo maps of \ion{H}{1} emissions \citep{sch98}. Also $E(B-V) \sim 0.2$ is  
obtained from the dust map based on analysis of 2MASS photometry \citep{mar06}.

\subsection{Distance Estimates from Pulsating RG}\label{subsec_IRmag}

An independent way of distance estimate comes from the pulsating RG companion of 
PU Vul. There is a well-known relation between the pulsation period and its luminosity 
for Mira variables, applicable also to semi-regular variables pulsating 
in the fundamental mode, 
\begin{equation}
M_K= -3.51 \times (\log P ({\rm day})- 2.38)-7.25, 
\label{equation_MkP}
\end{equation}
with an error of $\sim$ 0.2 mag \citep{whitelock08}. 
For the 218 day pulsation, we get the absolute $K$ magnitude of 
$M_K=-7.10 \pm 0.2$. 
During the 1980 eclipse the average $K$ mag changes from 6.26 mag to 6.51 mag 
\citep{bel85} and its average value is $<K>=6.37\pm 0.04$ mag. 

On the other hand, the difference between the absolute and apparent K-magnitudes is 
written as
\begin{equation}
K-M_K= 0.353 \times E(B-V) + 5 \log ~(d/1~{\rm kpc}) +10, 
\label{equation_distmod_IR}
\end{equation}
here we adopt the reddening law of $A_{\rm K}=0.353 \times E(B-V)$ \citep{cmm89}. 
Therefore, we get 
\begin{equation}
13.47= 0.353 \times E(B-V) + 5 \log ~(d/1~{\rm kpc}) +10.
\label{equation_distmod_IR2}
\end{equation}
This relation is plotted in Figure \ref{pu_vul_extinction}. 
For a particular value of $E(B-V)=0.3$, we get $d=4.7$ kpc. 
Figure \ref{pu_vul_extinction} shows that the intersections of the three solid 
lines, i.e., UV, optical, and IR, are relatively close, and $E(B-V)=0.3$ from  
optical color excess is also close to these points. This indicates that our various 
methods are consistent with each other. Considering ambiguity of each method, we use 
$E(B-V)=0.3$ and $d=4.7$ kpc as a reasonable estimate.

\placetable{table_TandRofWD}
\begin{deluxetable*}{lllllllllll}
\tabletypesize{\scriptsize}
\tablecaption{Temperature and Radius of Hot Component\tablenotemark{a}
\label{table_TandRofWD}}
\tablewidth{0pt}
\tablehead{
\colhead{Date} &
\colhead{JD} &
\colhead{Spectral} &
\colhead{Ion\tablenotemark{b}} &
\colhead{$V$} &
\colhead{$T_{\rm h}$} &
\colhead{$M_{\rm bol}$} &
\colhead{$L_{\rm h}$} &
\colhead{$R_{\rm h}$} &
\colhead{Method\tablenotemark{c}}\\
\colhead{}&
\colhead{2,400,000+}&
\colhead{type}&
\colhead{}&
\colhead{}&
\colhead{(K)}&
\colhead{}&
\colhead{($L_\odot$)}&
\colhead{$(R_\odot$)}&
\colhead{}
}
\startdata
04/1979   & 43980 & A7 II & & 8.84   &  7900  &  -5.36 &  11070 &  56  & [1]\\
05/1979  & 43991 & F0 I & & 8.93   &  7400  &  -5.23 &  9820 &  60  & [1]\\
07/1979  & 44070 & F0 I & & 8.84   &  7400  &  -5.32 &  10670 &  63  & [1]\\
12/1979   & 44222 &  F3 I& & 8.78 &    6900 & -5.41  &  11590 &  75  & [1]\\
08/1981 & 44834 & F5 I  & & 8.55  &  6500  & -5.66    &   14590 &    95 & [1]\\
09/1981 & 44873 & F5 I  & & 8.50 & 6500 & -5.65 & 14510 &  95 & [2]\\
06/1982 & 45147 & F5 I & & 8.50 & 6500 & -5.71 & 15280 & 97 & [1]\\
09/1982 &  45229 & F5-8 I & & 8.67 & 6300 &  -5.57   &  13430  & 97  &[1] \\
11/1982  & 45290 & F3-4 I & & 8.40  & 6800 &  -5.79   & 16440  &  92 &  [1]\\
12/1982   & 45320 &  F0 I & & 8.43 & 7400  &  -5.73    & 15560  & 76  & [1]\\
07/1983 &   45533  & F0 I & & 8.39  &7400 & -5.77   & 16140 & 77  & [1]\\
10/1984 &   45989  & A3 I  & &8.55  &8900 &   -5.77    &  16140    &  53  & [1]\\
06/1985 &   46232 &  A3 I & & 8.52 &8900  &  -5.80  &    16600  &  54 & [1] \\
09/1986  &  46690 &  A2 I  & & 8.72 & 9200  &  -5.67   &  14720  &  48 &  [1]\\
01/1988  &  47176  & --  & S+   & --  & 10000\tablenotemark{d} &  --   &   -- & --  & -- \\
06/1988  &  47328  & -- & N+    &--  & 15000\tablenotemark{d}  &     --   & --    &    --  & -- \\
10/1988 &   47438 & --  & N+2  & --  & 35000  &  -5.43 & 11800  & 3.0 & [3] \\
10/1988 &   47438 &  -- & N+2  & --  & 29000  &  -5.49 & 10704  & 4.4 & [3a]\\
05/1989  & 47666 &  -- &C+3,N+3&  --& 40000 &  -5.47 & 12230 & 2.3 & [3] \\
05/1989  & 47666 & -- & C+3,N+3&  --& 48000 &  -5.47 & 15440 & 1.8 & [3a]\\
05-09/1989 & 47730 & -- & C+3,N+3&  --&  48000&   -5.70 &   15140  & 1.8  & [4]\\
04/1990 & 47795 & -- & O+3 & -- & 55000 & -5.81 &16750 & 1.4 & [4]\\
07/1990 &  48088& -- &  Ne+2&  -- &     41000\tablenotemark{d} & -5.64 & 14320 &2.4  & [4] \\
11/1990 &  48217& --  & He+2 &   --  &    65000 &    -5.65&   14500 & 0.95 & [5], SWP40155\\
04/1991 &  48352& --  & He+2 &  --  &    65000 &    -5.64&   14350 & 0.94 &  [5], SWP41299\\
09/1991 &  48522& --  & He+2 &  --  &    67000  &   -5.59&   13710 & 0.87 & [5], SWP42536\\
10/1991 &  48559& --  & He+2 &   -- &    70000  &   -5.55&   13230 & 0.78 &  [5], SWP42937/8\\
08/1992 &  48858& --  & He+2 &    --&     77000 &    -5.67 &  14660 & 0.68 & [5], SWP45415\\
06/1995 &  49886 & --  &  Ne+4 & -- & 97000 & -5.58 & 13540 &  0.41 & [6] \\
06/1995 & 49886  & --  & Ne+4 & -- &  97000 & -5.68 &  14860 &  0.43 &  [4] \\
06/1996 &  50237&  --& He+2 &  --  &     90000 &    -5.31&  10570 &   0.42 & [5], SWP57322/3\\
06/1996 & 50237 &   -- &  -- & -- &  90000 &   -5.42 &  11700 &   0.36 & [4]\\
09/1996 &  50342 & -- & He+2 &    -- &    83000 &    -5.05&   8300&    0.44& [5], SWP58251/2\\
09/2001 & 52154 &-- & Fe+6 & --& 100000 & -4.68 & 5940 & 0.26 & [6] \\
09/2001 & 52154 & -- & Fe+6 & -- & 100000 & -5.42 & 11700 & 0.36& [4]\\
26/3/2003&  52818&-- &  Fe+6 & -- &   165000 & -4.97 & 7730 & 0.11 & [4] \\
9/4/2004 & 53105 &-- &  Fe+6 & --&   $> 99000$&  --   &   --&  --    &  --   \\
04/2004 & 53110 & -- &-- &--  &  150000\tablenotemark{e} & -4.97 & 7720 & 0.13 & [4]\\
07/2006 & 53930 &-- & O+5 &--& $> 114000$ & -- & -- & -- & --\\
07/2006 & 53930 & -- & O+5 & -- &150000\tablenotemark{e} & -4.70 & 6030 & 0.11 & [4] \\
06/2011 & 55740 & -- & -- & -- & 150000\tablenotemark{e} & -5.02 & 8090 & 0.13 & [4]
\enddata
\tablenotetext{a}{We assume $d=4.7$ kpc and $E(B-V)=0.30$.}
\tablenotetext{b}{Highest ionization stage.}
\tablenotetext{c}{Methods used in deriving the results (see Section \ref{section_TR}
for details).
[1] Supergiant phase method;
[2] Integration of SED; 
[3] Black body fit to a short wavelength {\it IUE} spectrum with $T$ as a free parameter; 
and [3a] with $T$ from the highest ionization stage observed; 
[4] \citet{mue94} method based on $UB$ observations of the nebular phase;
[5] analysis of \ion{He}{2}1640 emission line and ultraviolet continuum; 
[6] based on \ion{He}{2} 4686 emission line flux.}
\tablenotetext{d}{Ionization stage and $T_{\rm h}$ taken from \citet{mue94}.}
\tablenotetext{e}{Arbitrary assumed.}
\end{deluxetable*}

\begin{figure}
\epsscale{1.15}
\plotone{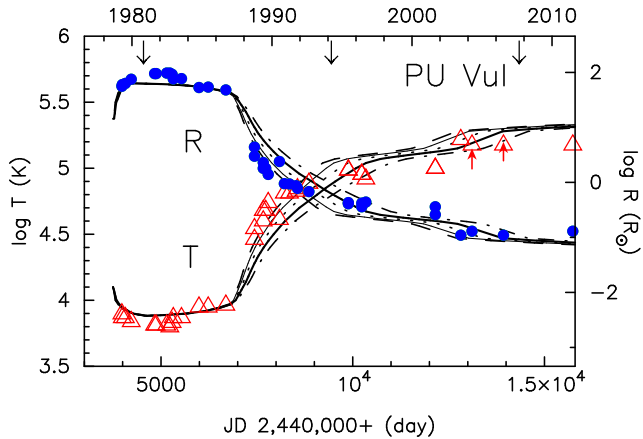}
\caption{Development of the temperature (open triangles and the left 
ordinates) and radius (filled circles and the right ordinates) of the hot 
component (WD) of PU Vul, taken from 
Table \ref{table_TandRofWD}. Two red upward arrows indicate lower limits  
of temperature. 
Model lines denote the photospheric temperatures and radii of the 0.6 $M_\odot$ 
WD with $X=0.7$ and $Z=0.01$ for five different wind mass-loss 
rates. 
Dash-dotted lines: $4 \times 10^{-7}~M_\odot~$yr$^{-1}$.  
Thick solid lines: $5 \times 10^{-7}~M_\odot~$yr$^{-1}$ (Model 1).  
Dotted lines: $6 \times 10^{-7}~M_\odot~$yr$^{-1}$.  
Thin solid lines: $7 \times 10^{-7}~M_\odot~$yr$^{-1}$. 
Dashed lines: $8 \times 10^{-7}~M_\odot~$yr$^{-1}$.
Three black downward arrows indicate the central times of the 1980, 
1994, and 2007 eclipses. 
\label{Temp}}
\end{figure}

\placefigure{Temp}

\section{Evolution of Temperature and Radius}\label{section_TR}

We have deduced the temperature and total luminosity of the 
hot component (WD) of PU Vul using {\it IUE} spectra as well as published photometry 
and other useful information. 
The radius is calculated from the temperature and luminosity using 
the Stefan-Boltzmann law. 
Here we assume $E(B-V)=0.30$ and $d=4.7$ kpc. 
We use the extinction law of \citet{fit07} 
for the narrow band continuum and emission lines, and \citet{cmm89} 
for the broad band optical and near infrared photometry, respectively. 
The resulting values are listed in  Table \ref{table_TandRofWD}. 
For several epochs we have been able to calculate them using more than one 
method, and the differences in the results give an idea about the uncertainties 
of these methods.

The third column of Table~\ref{table_TandRofWD} shows the spectral 
classification adopted from \citet{kol83}, \citet{bel84}, and \citet[][Table 3]{mue94}, 
and the fifth column shows the average of observed $V$ magnitudes \citep{kol83,bel82a,bel85,bel90} 
of PU Vul during the optical maximum, 1979--1986.
The outbursting component of PU Vul showed spectra similar to that of 
an early F-type supergiant, gradually evolving towards an A supergiant. 
Since there is no strong nebular contribution during this phase, 
we have assumed that the spectral type is an appropriate indicator 
for the effective temperature and bolometric correction, and 
that the observed $V$ magnitude \citep{kol83,bel82a,bel85,bel90} 
represents that of the hot component. 
We have adopted the temperatures and bolometric corrections for 
A-F supergiants from \citet{str81}, and adopt $M_\odot$(bol)=4.75 for the 
absolute bolometric luminosity of the Sun.

The hot component luminosity in September 1981 
(JD 2\,444\,873) have been estimated directly by integrating the spectral 
energy distribution (SED) from ultraviolet to infrared. To get the SED 
we have combined the very long exposure {\it IUE} spectra (SWP 15110, LWR 11627 
and LWR 11628) from 27 September 1981 with \citet{bel85} spectrophotometry 
performed on 25 September 1981 and $JHK$ photometry obtained on 
22 September 1981. The SED has been corrected for the reddening. 
The resultant bolometric magnitude is $m_{\rm bol,0}=7.71$, which derives 
the absolute bolometric magnitude $M_{\rm bol}=-5.65$ with the distance $d=4.7$ kpc. 
This value, $-5.65$, shows an excellent agreement with 
the average value, $-5.66$ in August 1981, derived from the observed $V$ mag and spectral 
type (see Table~\ref{table_TandRofWD}).

We have also used this SED to estimate the bolometric correction corresponding 
to this particular date, in September 1981. We obtain $V_0=7.57$ from the SED. 
Thus, the bolometric correction is calculated as 
BC$(V)=m_{\rm bol,0}- V_0$ = 0.14.  
On the other hand, \citet{bel85} obtained $V=8.45$ on 
JD 2\,444\,873, which is corrected to be $V_0=7.52$ with an extinction of 
$3.1 \times E(B-V)$=0.93. Combining this with $m_{\rm bol,0}=7.71$, 
we get BC$(V)=0.19$. These BC$(V)$ values are somewhat larger than 
BC$(V)=0.08$ corresponding to F5 I spectral type estimated at this epoch.  
The difference may reflect the lower density in the nova envelope 
than that in the brightest F-type supergiants.
In the present work, we use BC$(V)=0.17$, the mean value of 0.14 and 0.19.

For the nebular phase, whenever possible, the temperature of the hot 
component (WD), $T_{\rm h}$, has been estimated from the equivalent width 
of the \ion{He}{2} 1640 emission line measured from {\it IUE} spectra (identified 
in the last column of Table \ref{table_TandRofWD}). 
Although the high resolution {\it HST/GHRS} spectrum taken in October 1994 
(the 1993/4 eclipse egress) suggest $\sim 20\%$ of \ion{O}{1}]1641 line 
contribution to the \ion{He}{2} fluxes derived from 
lower resolution spectra, the \ion{O}{1}]1641 line is not visible in the 
well exposed high resolution spectra SWP45417 and SWP57730 taken 
before and after the eclipse, respectively.  
Therefore, we assume a negligible contribution of \ion{O}{1}] to our 
measurements of the \ion{He}{2} 1640  line. 
The luminosity of the hot component, $L_{\rm h}$, has been 
calculated from the \ion{He}{2} 1640 flux, assuming 
blackbody ($T_{\rm h}$)   
and that the \ion{He}{2} lines are produced by photoionization followed by 
recombination (case B). 
$L_{\rm h}$ has been also estimated from the 
{\it IUE} flux at 1350 ~\AA, assuming that it is emitted by 
blackbody ($T_{\rm h}$). At most epochs in Table \ref{table_TandRofWD} 
these two values agree with each other, 
and a mean value is adopted for the final $L_{\rm h}$.

At several epochs, $T_{\rm h}$ has been derived from the highest 
ionization potential (IP) observed in the {\it IUE} spectra 
(JD 2\,447\,438--795) 
and published optical spectra \citep{mun02,yoo07,tat11}.
We used the relation $T_{\rm h}$/1000 $\sim$ IP (eV) found by \citet{mue94}. 
For two epochs of JD 2\,452\,154 and 2\,452\,818,  $T_{\rm h}$
is derived from the ratios of H$\beta$, \ion{He}{2} 4686 and 
\ion{He}{1} 5876 emission line fluxes published by 
\citet{tat11} and \citet{lun05}, respectively.

At two epochs of 1988 and 1989, the hot component
parameters have been derived by fitting a blackbody to the short wavelength
($\lambda \la 1590$ ~\AA) part of the spectrum obtained by combining SWP34405, SWP34406, 
and SWP34407 for JD 2\,447\,438,  and SWP36301, SWP36302 and SWP36304 for JD 2\,447\,666.

After 1996, in the absence of {\it IUE} spectra, $ L_{\rm h}$ has been derived 
from the $UB$ mag observed by \citet{bel00} and by \citet{shu11}  
with the method proposed by \citet{mue94}. 
This method assumes that after subtraction of the 
contribution from the RG, the optical magnitudes contain a direct
contribution from the hot star and an indirect contribution from the nebulae. 
Thus, an accurate estimate of the RG contribution is especially important. 
The RG companion is classified to be a spectral type of M6 
(Section \ref{subsection_1st}), so its contribution to $UB$ magnitudes 
can be neglected.  
In fact, the $B-V$ and $U-B$ colors \citep{shu11} 
suggest that the continuum is still dominated by the nebular emission, 
in agreement with the optical spectra showing only faint flat continuum and strong 
emission lines \citep[e.g.][]{yoo07}.
As described later (Sections \ref{sec_eclipse} and \ref{sec_opticallightcuve} 
and Figure \ref{light}), 
the summation of the WD and nebular contributions dominates the $V$ light curve, 
although it shows a clear $\sim 0.5$ mag pulsation owing to the RG. 
Therefore, we can safely use the method of \citet{mue94}.
\citet{mue94} also provided the bolometric corrections to $UBV$ mag of
hot component for a wide range of hot component temperature,
$T_{\rm h}$. These bolometric corrections
were derived by model calculations with hot component temperature and
nebular density as free parameters.
Although the RG in PU Vul is similar to a Mira component of 
D-type symbiotic systems, we have decided to use the bolometric corrections 
for S-types because the $U-B$ color of PU Vul during the nebular phase 
is similar to that predicted by the \citet{mue94} 
model for S-types, 
and the electron density derived for the nebular phase is similar 
to the values characterizing the other S-type systems \citep{lun05}.

At JD 2\,449\,886 and JD 2\,452\,154,
the luminosity of the hot component, $L_{\rm h}$ has been calculated
from the published \ion{He}{2} 4686 fluxes \citep{and95,tat11},
assuming a blackbody spectrum
with $T_{\rm h}$ and that the \ion{He}{2} lines are produced by
photoionization followed by recombination (case B).

Our estimated temperature and radius are plotted in Figure \ref{Temp}. 
This figure also shows theoretical models for the 0.6 $M_\odot$ WD 
with the composition of $X=0.7$ and $Z=0.01$, i.e., 
the same models as in Figure \ref{lightUV4.f}. 
The higher the mass-loss rate, the faster the evolution. 
All of these models show more or less good agreement with 
our observational estimates. 
It should be noticed that the theoretical values are those of a blackbody 
photosphere. Even though, they show good agreement with the {\it IUE} flux 
(Figure \ref{lightUV4.f}) and also with the estimates 
obtained with quite different methods (Table \ref{table_TandRofWD}). 

\begin{figure}
\epsscale{1.15}
\plotone{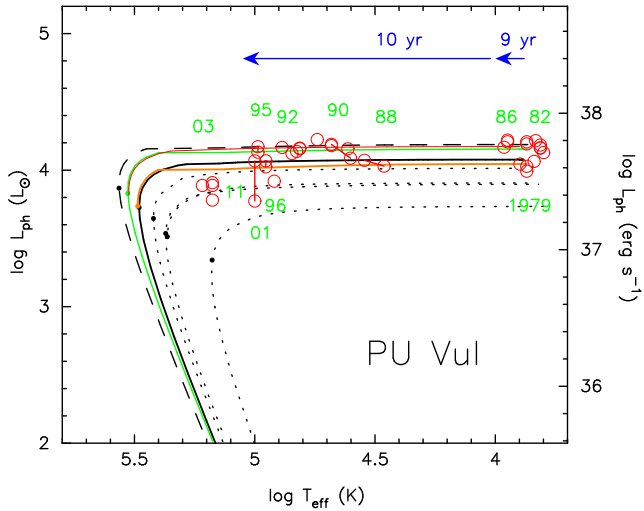}
\caption{
Evolution of the hot component of PU Vul in the HR diagram. Red open circles are our 
observational estimates, taken from in Table \ref{table_TandRofWD}. 
Pairs connected by a line segment indicate the estimates obtained for the same 
day data but with different methods. 
Observational years (two digits except 1979) are attached beside the point. 
Various types of lines denote loci of theoretical models. 
Solid lines indicate, from upper to 
lower, Model 4 (red), Model 3 (green), Model 1 (black), and Model 2 (brown). 
A dashed line denotes a 0.7 $M_\odot$ WD with $X=0.7$ and $Z=0.01$. 
Four dotted lines denote 0.5 $M_\odot$ WD models with different 
radius and chemical compositions. 
From upper to lower, a cold WD (the Chandrasekhar radius) with 
$X=0.5$ and $Z=0.02$, a cold WD with  $X=0.7$ and $Z=0.01$ (upper) which is 
almost overlapped with a cold WD with $X=0.7$ and $Z=0.02$ (lower). 
The lowest line denotes  a hot WD with $X=0.7$ and $Z=0.02$. 
Dots represent the epoch when nuclear burning extinguishes. 
Arrows indicate evolution timescales of Model 1; from the beginning to 
$\log T_{\rm ph}$ (K)=4.0, and from $\log T_{\rm ph}$ (K)=4.0 to 5.05, 
i.e., 8.7 yr and 9.6 yr, respectively.
\label{HR}}
\end{figure}

Figure \ref{HR} shows the HR diagram of the hot component. 
This figure shows theoretical tracks of the 0.5, 0.6 and 
0.7 $M_\odot$ WDs for various chemical compositions. 
The observational estimates are also in good agreement with our 
0.6 $M_\odot$ models.

\placefigure{HR}

\section{Light Curve Models of Eclipses} \label{sec_eclipse}

Now we present light curve models of the first (1980) and second (1994) eclipses.  
Here we assume spherical shapes of the both components,  
that the inclination angle of the orbit is $i=90 ^\circ$, that the RG 
moves in front of the hot component (WD) with a constant velocity $V_{\rm orb}$,  
and that the RG is radially pulsating with a period of 218 days  and  
its flux changes in a sinusoidal shape around the equilibrium magnitude.  
We also assume that the radius of the RG also varies 
in a fashion of long-period Mira variables
\citep{tho02,woo04,woo08}, and that the radius varies 
sinusoidally with a phase shift of 0.5 to the flux variation, i.e., 
the radius reaches the minimum at the maximum brightness as reported by 
\citet{shu11}. 
No accretion disk is assumed, because there is no observational 
indication.

\begin{figure}
\epsscale{1.15}
\plotone{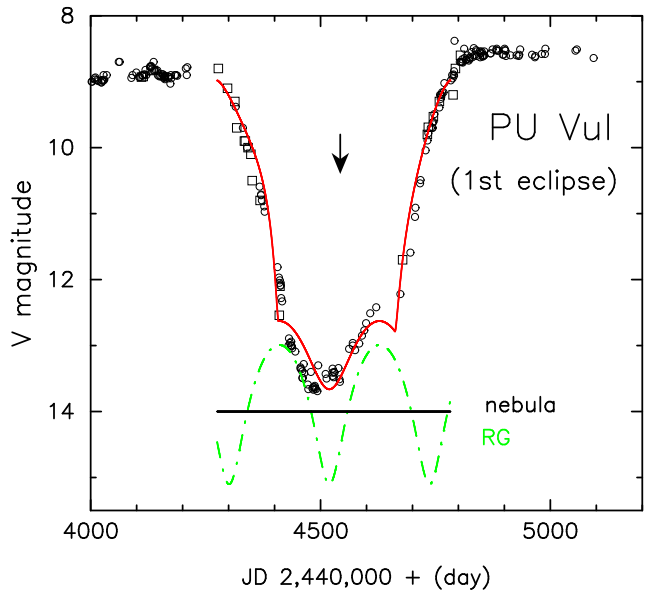}
\caption{A close up view of the first eclipse. 
Our model light curve is indicated by the red solid line, which is 
a summation of the eclipsed WD photosphere, a constant nebular 
emission of $V=14.0$, and the RG photosphere with a sinusoidal oscillation 
around $V=13.6$ (green dash-dotted line).
The mideclipse on JD 2,444,532 is indicated by a downward arrow. 
See text for more details.
}
\label{eclipse.1st}
\end{figure}

\placefigure{eclipse.1st}

\placetable{table_eclipse} 
\begin{deluxetable*}{llllll}
\tabletypesize{\scriptsize}
\tablecaption{Eclipse Light Curve Model
\label{table_eclipse}}
\tablewidth{0pt}
\tablehead{
\colhead{Subject} &
\colhead{} &
\colhead{1st eclipse} &
\colhead{2nd eclipse} &
\colhead{Units}
}
\startdata
mideclipse& ...           &4,532  &9,447& JD 2,440,000+  \\
total duration (D)& ...    &508  &345& day  \\
totality (d)    & ...     &254  &343& day  \\
mean RG magnitude & ...   &13.6  &13.6& mag  \\
amplitude of RG luminosity & ... &75 \% &65 \% & \\
total amplitude of RG in mag\tablenotemark{a} & ... &2.1  &1.7&mag\\
amplitude of RG radius & ...&7 \%& 3 \%& \\
$R_{\rm c}/a$ & ...&0.246& 0.22&  \\
$R_{\rm h}/a$ & ...&0.070& 0.0007 &  \\
nebular emission  & ...&14.0& see Fig. \ref{eclipse.2nd}& mag
\enddata
\tablenotetext{a}{$V$(min)-$V$(max)}
\end{deluxetable*}

\subsection{The First (1980) Eclipse} \label{subsection_1st}

We suppose that the 1980 eclipse is total, in which the bloated WD 
is completely occulted by the pulsating RG companion.
The bottom magnitude of $V \sim 13$ during the eclipse seems to be a bit 
higher than that of a late type M-giant, which suggests the presence 
of a weak emission source 
which was not occulted. Before going to our model construction, we 
need to examine the magnitude of the M-giant companion. 

The spectral classification of the RG companion is estimated 
to be M3--M7, but better estimates are obtained in longer wavelength bands  
rather than in the optical because of contamination by nebular emission. 
\citet{ms99} obtained M6--7, using the bands in near IR, i.e. $\lambda 
\gtrsim 8000$ ~\AA. 
\citet{bel85} derived a similar type, M6.5, during the 1980 eclipse. 
This value is uncorrected for the faint nebula ($V\sim 14$), so there may be 
some fluctuations by $\pm 1$ in the spectral type. 
Therefore, we regard M6 as a reasonable average spectral type of the M-giant.

The magnitude of the RG can be estimated from its $K$-band magnitude. 
\citet{bel85} obtained $<K>=6.37$ during the eclipse, and its reddening 
corrected value is $<K>_0=6.26$ (see Section \ref{subsec_IRmag}).  
A similar value is obtained $<K>_0=6.17 \pm 0.01$ from \citet{bel00} and \citet{tat11} 
for an average over 1989--2009. 
As $V-K\sim 7.2$ for an M6 III star, \citep[e.g.][]{str92}, we get $<V>_{\rm 0}\sim 13.4$.
Assuming $E(B-V)=0.3$, the visual magnitude of the giant becomes  
$<V> \sim 14.3$. This value is consistent with an averaged pre-outburst 
magnitude of $V=14.1 \pm 0.15$ \citep{step79} and $B=16.5$ \citep{lil79}. 
This magnitude $<V> \sim 14.3$ is much darker than the observed mean 
magnitude of $V \sim 13$ at the bottom of the eclipse, so we need an additional 
source of emission possibly originated from optically-thin plasma such as 
heated RG cool winds. 

We have constructed an eclipse light curve model, assuming that the RG mean 
magnitude, amplitude of the pulsation, and brightness of the additional emission 
source are free parameters.   
Figure \ref{eclipse.1st} shows a close-up view of the first eclipse and 
our light curve model.

A model light curve, that produces a better fit to the observed magnitude 
data, is shown in Figure \ref{eclipse.1st}.
We obtain the total duration of the eclipse $D = 508$ days, the totality $d = 254$ days,  
and $R_{\rm c}^*/a=(D+d)/(2\pi P_{\rm orb})=0.244$, where $a$ is the separation of 
the two stars, $P_{\rm orb}$ is the orbital period in units of day, and  
the asterisk denotes the specific radius,  
because it depends on the timing of pulsations at the ingress/egress. 
The equilibrium radius of the pulsating RG is $R_{\rm c}=0.247~a$.
The RG radius is smaller than the equilibrium radius at 
the second and third contacts, i.e., 0.93 and 0.96 times the equilibrium radius, 
we obtain $R_{\rm c}/a$ slightly larger than $R_{\rm c}^*/a$. 

A bottom magnitude is obtained as a combination of the RG equilibrium magnitude and 
the nebular emission. An equilibrium magnitude of the RG darker than $V = 13.8$ does not 
reproduce the wavy bottom shape, even if we assume a very large amplitude of 
the luminosity. A combination of the RG equilibrium magnitude of $V=13.6$ -- 13.8 and 
a nebular emission of $V \sim 14.0$ yield a better fitting 
as shown in Figure \ref{eclipse.1st}. Here we assume the RG equilibrium magnitude 
to be $V=13.6$ and the nebular emission $V=14.0$.

For the oscillation of the RG radius,  
a good fitting is obtained for amplitudes of the radius $\Delta R_{\rm c}/R_{\rm c}=
$ 0.06--0.08. 
For a larger amplitude $\Delta R_{\rm c}/R_{\rm c} \gtrsim 0.1$,
we cannot find a good shape of light curves, because 
a wavy structure appears during the ingress and egress.  
Here, we adopt $\Delta R_{\rm c}/R_{\rm c}=0.07$. 

Our fitting parameters are summarized in Table \ref{table_eclipse}, 
showing the mideclipse time, i.e., the time when the RG center comes just in front 
of the WD, the total duration of eclipse ($D$), 
totality ($d$), apparent magnitude of the RG at its equilibrium state,
amplitude of the RG luminosity in linear scale $\Delta L_V/L_V$, 
corresponding to the total amplitude in magnitude (difference between the 
maximum and minimum magnitudes), 
amplitude of the RG radius,   
the ratio of the RG radius to the separation $a$, and the ratio of the WD radius to $a$. 
Note that the RG magnitude at equilibrium is not the arithmetic mean of 
the maximum and minimum magnitudes, because we assume a sinusoidal variation in 
the luminosity (linear scale), not in the magnitude (logarithmic scale).

\citet{gar96} supposed that this eclipse is partial because of a non-flat 
bottom. \citet{vog92} explained this non-flat bottom shape 
as a total eclipse, but 
contaminated by two nebular emissions that cause the flux excess 
in the early half and later half, respectively. 
However, the wavy bottom shape in Figure \ref{eclipse.1st} 
is consistent with the RG pulsation. 

\begin{figure}
\epsscale{1.15}
\plotone{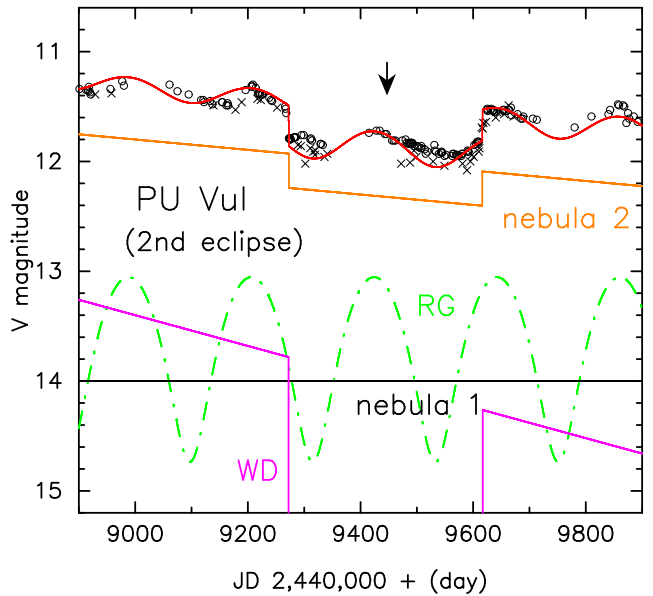}
\caption{A close up view of the second eclipse. 
Observational data are taken from \citet{kol95} (crosses) and \citet{yoo00} 
(open circles). 
The red solid line indicates our composite light curve which is 
a summation  of the four components, i.e., the pulsating RG 
(green dash-dotted line), WD (magenta solid line), constant ``nebula 1'' emission 
of $V=14.0$ (black solid line),
and gradually decreasing ``nebula 2'' emission which is eclipsed by 25 \% 
(brown solid line). 
The RG is pulsating around its mean magnitude of $V=13.6$.
The mideclipse on JD 2,449,447 is indicated by a downward arrow.}
\label{eclipse.2nd}
\end{figure}

\placefigure{eclipse.2nd}

\subsection{The Second (1994) Eclipse} \label{subsection_2nd}

PU Vul began to decline in 1987 from the flat maximum and  
reached $V\sim 11.5$ just before the second eclipse in 1994. 
The bottom magnitude of the second eclipse is $V\sim 11.8$ 
(see Figure \ref{eclipse.2nd}), 1.8 mag brighter than that of the 1980 eclipse. 
This eclipse is considered to be total, because the continuum UV radiation,  
which has a WD origin, was disappeared during the eclipse \citep{nus96,tat09}. 
Therefore, the excess flux ($\Delta V \sim 1.8$ mag) is a contribution of 
hot nebulae. 

\citet{nus96} analyzed UV spectra of the hot nebulae and found that 
highly ionized lines disappeared during the eclipse and recovered after 
that, whereas low-ionized nebular lines were hardly affected. 
This means that the high excitation lines 
were emitted from a region close to the WD and 
the low-ionized nebular lines are emitted from an outer extended region.
In other words, the nebulae are also partially occulted.

Thus, there are three sources of emission: the pulsating RG, 
totally occulted WD, and partially occulted nebulae.
For the RG, we assume a similar model as in the first eclipse, i.e.,  
the RG is pulsating around the equilibrium magnitude of $V=13.6$  
but maybe with different amplitudes of the luminosity and radius, which are 
fitting parameters. 
For the WD emission, we take Model 1, which is shown in Figure \ref{eclipse.2nd} 
(labeled as WD). 
We assume two sources of nebular emission, one is a constant component 
(labeled ``nebula 1'')   
assumed in the first eclipse, i.e., $V=14.0$. The other is a decreasing 
component which is partially occulted during the second eclipse (labeled  
``nebula 2''). Its luminosity and decline rate are also parameters 
in order to obtain the best fit. 

Figure \ref{eclipse.2nd} shows the resultant light curve. 
The amplitude of the luminosity is determined to be 65 \% and that 
of the radius is 3 \%. 
For the nebula 2 component, we found that a 25 \%  occultation of 
the nebula 2 emission yields the best fit. 
We see that our composite light curve represents the temporal 
change of the optical data.

These fitting parameters are summarized in Table \ref{table_eclipse}.  
It is difficult to obtain the WD radius from the light curve fitting 
because the ingress and egress of the second eclipse is very steep, 
which indicates the eclipsed object is very small. 
Therefore, we fixed the WD radius to be $R_{\rm h}/a=0.0007$, 
which corresponds to $1.0 R_\odot$ for 
a circular orbit of a binary consisting of a $1.0 M_\odot$ RG and a $0.6 M_\odot$ WD. 
This assumption has no effects in determining the other parameters.

\placefigure{eclipse.2nd3rd}
\begin{figure}
\epsscale{1.15}
\plotone{f7.epsi}
\caption{Light curve of PU Vul for 
 the period JD 2,447,000--2,451,000 (upper part) and 
JD 2,453,500--2,457,500 (lower part).
Observational data are taken from 
\citet{kol95}(crosses), \citet{yoo00}(small open circles), 
\citet{kle94}(squares), \citet{kan91b}(open stars), and  
\citet{iij89}(middle size open circles).
For the lower part, observational data are taken from AAVSO (dots) 
and All Sky Automated Survey (ASAS) (crosses).
Downward arrows indicate the central times of the eclipses, 
JD 2,449,447 and 2,454,362.
Short vertical lines indicate epochs of the pulsation maxima 
of the M giant assuming a period of 218 days.  
\label{eclipse.2nd3rd}}
\end{figure}

\subsection{M-giant Pulsation and 3rd Eclipse}\label{sec_pulsation}

Figure \ref{eclipse.2nd3rd} shows a periodic modulation of the $V$ magnitude,  
which becomes prominent in the later phase of the outburst where the 
WD component becomes dark. 
This modulation is unclear in the flat maximum except 
the first eclipse (Figures \ref{eclipse.1st}), 
because the hot component is dominant. We regard that this modulation is 
caused by a pulsation of the RG.  

We obtained the pulsation period to be 218 days,  
assuming that the period is unchanged from the first eclipse
until 2010. This 218 day period can reproduce well both 
the first and second eclipses as shown in 
Figures \ref{eclipse.1st} and  \ref{eclipse.2nd}. 
\citet{cho98} obtained a 217 day period 
and \citet{shu11} obtained a 217.7 day period. Our value is consistent 
with these periods.

Figure \ref{eclipse.2nd3rd} shows that one of the minima of the RG pulsation 
accidentally coincides with the time expected for the third eclipse in 2007  
indicated by an arrow. 
This narrow dip is not the third eclipse of the WD, because the duration is too   
short, and the WD had already become very dark in the optical band 
(see Figure \ref{light}), and an occultation of the WD hardly changes 
the total brightness of PU Vul. 
\citet{shu11} showed that the $U$ magnitude is clearly eclipsed in the third eclipse, 
but the $V$ magnitude is not. Our interpretation is consistent with theirs.

\placetable{table_RGradius}
\begin{deluxetable}{lllllll}
\tabletypesize{\scriptsize}
\tablecaption{Radii of the Cool and Hot Components\tablenotemark{a}
\label{table_RGradius}}
\tablewidth{0pt}
\tablehead{
\colhead{RG mass} &
\colhead{} &
\colhead{ $a$} &
\colhead{$R_{\rm c}$(1st)\tablenotemark{b}} &
\colhead{$R_{\rm c}$(2nd)\tablenotemark{c}} &
\colhead{$R_{\rm h}$(1st)}\\
\colhead{$(M_\odot$)} &
\colhead{}&
\colhead{($R_\odot$)} &
\colhead{($R_\odot$)}&
\colhead{($R_\odot$)}&
\colhead{($R_\odot$)}
}
\startdata
3.0 & ...    &1860  &459 &413 &131  \\
2.0 & ...    &1670  &411 &370 &118  \\
1.5 & ...    &1550  &383 &345 &109  \\
1.0 & ...    &1420  &350 &315& 100\\
0.8 & ...    &1360  &335 &301& 95.6 \\
0.6 & ...    &1290  &318 &286 &90.9 \\
0.4 & ...    &1210  &299 &269 & 85.5
\enddata
\tablenotetext{a}{A circular orbit and a 0.6 $M_\odot$ WD are assumed. }
\tablenotetext{b}{(1st) means the value obtained for the 1st eclipse}
\tablenotetext{c}{(2nd) from the 2nd eclipse}
\end{deluxetable}

\subsection{Radii of Cool/Hot Components}\label{subsection_radius}

In Sections \ref{subsection_1st} and \ref{subsection_2nd} 
we have obtained  $R_{\rm c}/a$ and $R_{\rm h}/a$ 
for the first and second eclipses.  
Assuming a circular orbit, we calculated   
the binary separation $a$ from Kepler's third law, 
$a^3 =G(P_{\rm orb}/2 \pi )^2(M_{\rm WD} + M_{\rm RG})$. 
The resultant radii of the RG and WD  
are listed in Table \ref{table_RGradius} as well as $a$ 
for various RG masses. Here we assume a 0.6 $M_\odot$ WD mass. 
Estimated RG radii (270 -- 460 $R_\odot$) seem to be a little bit 
larger than those of low mass M giant stars, which will be discussed 
in Section \ref{sec_discussion} (Discussion).

For the hot component, we obtain a radius of $R_{\rm h}/a=0.07$ for 
the first eclipse, that corresponds to $R_{\rm h} \sim$ 85--130 $R_\odot$ as shown in 
Table \ref{table_RGradius}.  
For the second eclipse, we have fixed the WD radius to be 
$R_{\rm h}/a=0.0007$, because the ingress and egress are too steep 
to determine the radius. 
Thus, it is not listed in the table. The steep decline/rise suggest 
$R_{\rm h}/a < 0.001$ which corresponds to $R_{\rm h} <$ 1--2 $R_\odot$. 
We can only say that the radius of the hot component reduced by a factor of 100 
or more.

In Section \ref{section_TR} we have already shown  that 
the WD photosphere had shrunk by about two orders of magnitudes between 
the first and second eclipses, from both the observational estimates 
and theoretical models (see Figure \ref{Temp}). 
The above radius estimates from the eclipses are very consistent with these 
estimates in Figure \ref{Temp}. 
This is the first time that the shrinkage of a nova WD photosphere has 
been measured by eclipse analysis.

\begin{figure*}
\epsscale{1.0}
\plotone{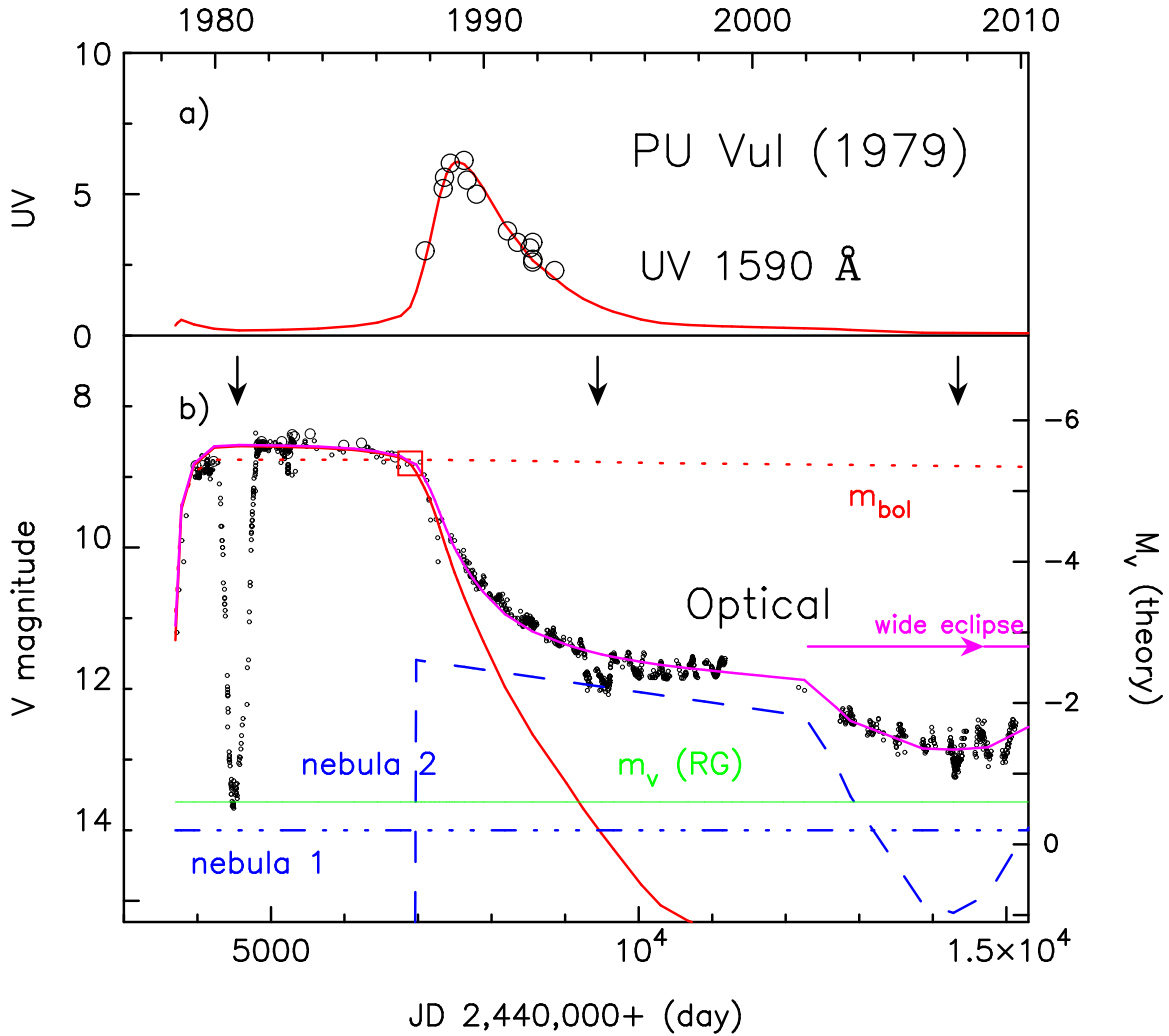}
\caption{Optical and UV light curves of PU Vul.
(a) UV light curve. 
Large open circles denote the {\it IUE} UV 1590~\AA~ band (the same as  
those in Figure \ref {lightUV4.f}d). The solid curve denotes the UV light curve 
of Model 1. The scale is in units of 
$10^{-13}$ erg~s$^{-1}$~cm$^{-2}$~\AA$^{-1}$.
(b) Optical light curve. 
Middle size open circles are $V$ magnitudes during the flat phase listed in Table 
\ref{table_TandRofWD}. See \citet{kat11} for the other observational data.  
The red solid and dotted lines indicate optical and bolometric light 
curves of Model 1.  
The large open square indicates the epoch at $\log T_{\rm ph}$(K)=4.0  
when the optically thin wind of $5 \times 10^{-7}$ $M_\odot$~yr$^{-1}$  
started in our model. 
The magenta solid line indicates a composite light curve of the WD (red solid 
line), RG at the  mean luminosity $V=13.6$ (horizontal green solid line), 
emission of a constant component at $V=14.0$ 
(``nebular 1'': blue dash-three-dotted line), 
and a variable component (``nebular 2'': blue dashed line). 
Downward arrows indicate the mideclipses. 
We suppose that the ``nebula 2'' is eclipsed by the RG companion after 2002 
(denoted by 'wide eclipse').
\label{light}}
\end{figure*}

\placefigure{light}

\section{Composite Optical Light Curve}  \label{sec_opticallightcuve}

Figure \ref{light} shows an observational light curve of PU Vul from the beginning
of the outburst until 2010.  It also depicts our theoretical composite
light curve that consists of three components, the WD, RG, and nebulae.  
The WD component, in which we use Model 1, well reproduces the observed 
UV light curve (Figure \ref{light}a) as well as the optical light curve until 1989.
After 1989, PU Vul entered a nebular phase and emission-lines dominate 
the spectra \citep{iij89, kan91b}.
Our WD model does not include line-emission formed outside the photosphere,
thus the $V$-light curve (red solid line) decays much faster than the observed one. 
For the RG component, we assume that the equilibrium magnitude is 
constant, $V=13.6$, throughout the outburst as we did in the first and 
second eclipses in Section \ref{sec_eclipse}. 
This $V=13.6$ is indicated by the green horizontal solid line
in Figure \ref{light}b. 

For the nebular emission, we assume two components: one is a constant component 
of $V=14.0$ as depicted by the dash-three-dotted line (nebula 1) in Figure \ref{light}b. 
We assumed this component uneclipsed at all during 
the first and second eclipses as shown 
in Figures \ref{eclipse.1st} and \ref{eclipse.2nd}.
We suppose this nebula 1 emission originated from the RG cool wind, partially  
ionized by the radiation from the hot component.
As this emission is faint, it dominated the total magnitude only in the 
first eclipse, so we have no information on its magnitude whether it changed or not. 
Therefore, we assumed that this component is constant. 
\citet{tat11} found the Raman scattered \ion{O}{6} 6830 line in the 
optical spectra taken in mid 2006 and later. 
This indicates the presence of neutral hydrogen, i.e., the RG cool wind. 
As the WD is still hot, a part of the RG cool wind may be ionized. 
So we reasonably suppose this component is still present. 

The other nebula is originated from
the WD, the shape of which is represented by the blue dashed line
(nebula 2).  This component
started at the epoch when the photospheric temperature of the WD increased to
$\log T$ (K)=4.0 (open square in Figure \ref{light}).
This WD-origin component was first discussed by \citet{nus88}, who 
concluded that the nebular emission is of WD-origin because the relative abundances 
of C, N, and O are close to those of classical novae but different from 
symbiotic stars. 
This component was eclipsed during the second eclipse as in Figure \ref{eclipse.2nd}. 

Recently, \citet{shu11} reported that all of the $U$, $B$, and $V$ magnitudes
are gradually rising after the third eclipse while the mean value of
the $I$ magnitude is almost constant.  
This indicates that the WD-origin nebular component is relatively
centrally condensed around the WD and, at the same time,
widely spread out over the orbit.
Then the width of eclipse by the RG is so wide that a whole period 
of the orbital phase is partially eclipsed 
as shown by the blue dashed line in Figure \ref{light}.
This increase in the brightness
($U$, $B$, and $V$) also suggests that the hydrogen shell-burning on the WD
is still on-going.

\section{Discussions} \label{sec_discussion}

\subsection{Comparison with Other Works on Eclipses} \label{subsec_comparison}

\citet{gar96} estimated the relative size of the cool component to be 
$R_{\rm c}/a=0.28$ for the first eclipse and $0.22$ for the second eclipse,   
assuming symmetric shapes of the eclipses. 
Our corresponding values are $R_{\rm c}^*/a=0.24$ and 0.22, respectively. 
The difference in the first eclipse is explained from the difference in the totality. 
Assuming that the bottom base line at the first eclipse was $V=11.8$ 
from the second eclipse, Garnavich get a larger totality than ours. 
Thus, $R_{\rm c}/a$ becomes larger than ours. 
In the second eclipse our $R_{\rm c}/a$ is essentially the same as 
Garnavich's, because the radius oscillation of the RG has little effects 
due to small amplitude (3 \%).
 
For the hot component, \citet{gar96} estimated $R_{\rm h}/a=0.1$ 
for the first eclipse, which is consistent with our value of $R_{\rm h}/a=0.07$, 
considering difficulty of accurate fitting with the scattered data. 
For the second eclipse Garnavich's value  $R_{\rm h}/a=0.02$ is much larger 
than our value of $R_{\rm h}/a = 0.0007$. This difference comes  
mainly from the different data sets. Garnavich used the AAVSO data 
that show slower decline/increase at the ingress/egress 
than those in Figure \ref{eclipse.2nd}. These AAVSO data, however, can be 
also fitted by a more steep light curve that yields $R_{\rm h}/a < 0.01$. 
Therefore, in the both cases, we can say that the radius of the hot component 
had decreased at least 
by a factor of ten between the first and second eclipses.

\citet{gar96} concluded that the RG radius shrunk  by 21 \% between the first and 
second eclipses, i.e., from $R_{\rm c}/a=0.28$ to 0.22. Our 
values are much smaller, 10 \% (from $R_{\rm c}/a= 0.246$ to 0.22), 
but the shrinkage of the radius seems to be real because  
we cannot find a parameter set for the same RG radius 
between the two eclipses. We will discuss the shrinkage of the radius 
in the next subsection.

The orbital period of PU Vul is estimated from the mideclipses of the first 
and second 
eclipses to be 4915 days (13.46 yr) (see Table \ref{table_eclipse}). 
The orbital period was obtained as $4918 \pm 8$ days \citep{kol95}, $4900 \pm 100$ 
days \citep{nus96}, $4900 \pm 9$ \citep{gar96}, 4897 days \citep{shu11}, 
assuming symmetric shapes of the eclipses. 
Our analysis first includes a radius oscillation and the resulted 
non-symmetric shapes of eclipses. However, these effects cause only 
several days off from the symmetry center because of 
small amplitudes of the radial oscillations. 
This is the reason why our new period is close to the previous estimates.

\subsection{Comparison with Other Evolution Calculations}

Figure \ref{HR} shows the evolution timescale of Model 1,
18.3 yr from the beginning of 
the outburst to $\log T_{\rm ph}$ (K)=5.05. 
If we do not include the optically-thin wind mass-loss, 
this becomes 46 yr, and the total duration of the outburst,  
from the beginning to the extinguish point of nuclear burning, 
is 130 yr. 
Using a hydrodynamical code  \citet{pri95} calculated 
multicycle nova evolution models for various 
WD masses and accretion rates.  
For a $0.65~M_\odot$ WD and a mass accretion rate
of $1 \times 10^{-7}M_\odot$yr$^{-1}$, no optically thick wind mass-loss 
arose. They obtained the total duration of the nova outburst to be 
$t_{\rm 3bol}$=155--176 yr, depending on the core temperature, where 
$t_{\rm 3bol}$ is the time during which the bolometric luminosity drops 
by 3 mag.  
For lower accretion rates ($\leq 10^{-8}M_\odot$yr$^{-1}$) strong optically 
thick winds occur which shorten the total duration. 
Their total duration is very consistent with our $0.6~M_\odot$ WD model, 
considering the different definition of the end point of a 
shell flash; Our definition is for the hydrogen burning extinguish point 
(depicted by the dot in Figure \ref{HR}), whereas
Prialnik \& Kovetz' is for $t_{\rm 3bol}$ time 
which comes later than our extinguish point, thus gives a longer timescale  
than ours.

Following the referee's suggestion we discuss the work 
by \citet{cas98} and \citet{pie99,pie00} who calculated  
shell flashes on low-mass WDs using  
a spherical symmetric hydrostatic code with the Los Alamos opacity.
Cassisi et al.'s (1998) models show that the envelope extends only down to
$\log T_{\rm ph}$ (K)=4.5--4.7 for a 0.5 $M_{\odot}$ WD  
with mass accretion rates of 2 and 4 $\times 10^{-8}M_{\odot}$yr$^{-1}$. 
Also in \citet{pie00}, the temperature decreases down to $\log T_{\rm ph}$ (K)
=4.1--4.2 only in a few exceptional cases. 
Such high-temperature shell flashes may be observed as a UV flash. 
In other words, these calculations do not represent realistic nova 
outbursts in which the surface temperature drops 
to $\log T_{\rm ph}$ (K)$~<~4.0$ at the optical peak. 
This suggests that their numerical code has some difficulties in calculating 
realistic nova outburst models. 

It should be pointed out that the above three works are obtained with the 
Los Alamos opacity, not with the OPAL opacity \citep{rog92,igl93,igl96}, 
which has been widely used in stellar evolution codes  
including nova outbursts. We are puzzled by the remark in Cassisi et al. 
that 
'the Los Alamos opacities are very similar to the OPAL opacities' 
\citep[see the last sentence of Section 4 in][]{cas98}. 
It is well known that the OPAL opacities have a strong peak
at $\log T$ (K) $\sim 5.2$, while the Los Alamos opacities do not 
\citep[For comparison with these opacities in a nova envelope, 
see Figure 15 in][]{kat94h}.
This strong peak causes substantial changes in nova outbursts, 
e.g., acceleration of optically thick winds,  and 
as a results, nova evolutions had significantly changed 
(e.g., compare hydrodynamical calculations of nova outbursts in 
\citet{pri86} obtained with the Los Alamos opacity
with \citet{pri95} with the OPAL opacity).
In a less massive WD ($\lesssim 0.6 M_\odot$), no optically thick wind is 
accelerated, but internal structures of the envelope are significantly 
different; a density inversion layer appears corresponding to the peak 
of the OPAL opacity \citep[see Figure 7 in][]{kat11}. 
In order to make a reliable outburst model of PU Vul, 
we need to use the OPAL opacity, not the Los Alamos opacity  
\citep[see also Discussion in][]{kat12}.

\subsection{Pulsating RG Companion}

As described in the previous subsection, 
our eclipse analysis shows that the RG radius decreased by $\sim$ 10 \% between 
the first and  second eclipses. This radius may not be the photospheric radius 
of the RG defined in near IR bands, but the radius of 
a thick TiO atmosphere, which is transparent in $K$-band  
but opaque in $V$-band. 
In the pulsating RG atmosphere, the temperature decreases 
in the expanding phase, which accelerates TiO molecule formation, resulting in 
a large opacity in the optical region, which causes a deep minimum in the optical 
light curve. 
The radius, that we obtained from the eclipses in $V$-band, 
corresponds to the radius of the TiO atmosphere. We call this the 
"visual photosphere" after \citet{rei02}.     
This radius could be much larger than the  
photospheric radius usually defined with $K$-band observation. 
Therefore, it is very likely that our visual photosphere in Table \ref{table_RGradius} 
is larger than the RG radius in $K$-band.

As shown in Section \ref{sec_pulsation} the pulsation period of 218 days had not 
changed between the first and second eclipses.
The unchanged pulsation period means that the internal structure 
of the RG had not changed, so the $K$-band photospheric radius should 
be the same. 
On the other hand, our analysis clarified that the pulsation 
amplitude in $V$-band decreased from 75 \% to 65 \%, 
and also the amplitude of the radius decreased    
(see Table \ref{table_eclipse}).
This suggests that the radius of the visual photosphere decreased as the 
amplitudes of the luminosity and radius had decreased. 
This can be understood as follows; The TiO atmosphere is pushed outward 
in an expanding phase, and it pushed far outward when its amplitude is larger.   
Therefore, a larger amplitude results in a larger visual photosphere.

We can estimate the RG radius, using the period-luminosity (PL) relations of 
Mira/semi-regular variables.   
The bolometric luminosity of LMC Mira variables follows a 
PL relation of 
\begin{equation}
 m_{\rm bol}=( -3.06 \pm 0.26)~\log P + (21.50\pm 0.61),
\label{equation_distmodLMC}
\end{equation}
where $P$ is the pulsation period in units of day \citep{gla03}.
The fundamental pulsation mode of Mira variables corresponds to this sequence. 
With the distance modulus of LMC, $18.39 \pm 0.05$ \citep{lee07}, 
we obtain $M_{\rm bol}=-4.05$ for $P=218$ day. 
Therefore, if the RG companion follows the PL relations of Miras, 
its absolute luminosity is $M_{\rm bol}=-4.05$, i.e., $\log L=3300~L_\odot$.
Photometric studies on a large number of stars indicate another PL relation, 
parallel to the above relation, 
but about one magnitude brighter. This sequence corresponds to the first overtone 
of pulsation in semi-regular variables.
In this case, we have $M_{\rm bol}=-5.05$, i.e., $\log L=830 0 L_\odot$.

The spectral type of the companion is estimated to be M6 (see Section 
\ref{subsection_1st}). 
The temperature calibration for late M giants is relatively well established, 
and various groups give similar values. In particular, \citet{rich99} give 
$T_{\rm eff}=3240 \pm 75$ K and $3100 \pm 80$ K for M6 and M7 giants, respectively, 
whereas \citet{vanbel99} report $3375 \pm 34$ K and $3095 \pm 29$ K for M6 and M7, 
respectively.
So, $T_{\rm eff}=3200 \pm 100$ K for the M6--7 giant in PU Vul seems very reasonable.
The radius then becomes  $R=187 \pm 12 R_\odot$. 
For the second PL relation we have $R=296 \pm 19 R_\odot$. 
Comparing these radii with the ones in Table \ref{table_RGradius}, we may 
conclude that the pulsation of the RG companion is consistent with 
the fundamental mode rather than the first overtone, because the 
visual photosphere is much larger than the stellar radius 
\citep[$R_{\rm c} \sim 1.8$ times the stellar radius, that is, 
$R_{\rm c} \sim 330~R_\odot$ for the fundamental mode:][]{rei02}.

If the RG pulsation is in the first overtone, the absolute magnitude is 
about 1 mag brighter than in the fundamental mode as described above, thus 
the distance is 1.6 times larger, i.e., $d=4.7$ kpc $\times 1.6 = 7.4$ kpc. 
Such a large distance is inconsistent with the optical light curve 
fittings, because $E(B-V)$ becomes 
too small or negative ( see black solid/dotted lines in 
Figure \ref{pu_vul_extinction}), thus we cannot 
construct a consistent model among the optical, UV 1590 ~\AA, and extinction.  
Note that the "IR" line in Figure \ref{pu_vul_extinction}, i.e., 
Equation (\ref{equation_MkP})  is for 
the fundamental mode and the corresponding line for the first overtone is in the 
right outside of the figure. Therefore, we may conclude that the pulsation is 
a fundamental mode.

Next, we estimate the RG mass using a theoretical relation obtained from 
radial-pulsations.  
It is well known that the pulsation constant, $Q=P{\sqrt{M/R^3}}$, is 
insensitive to stellar structure, here $M$ is the stellar mass in units of 
$M_\odot$, $R$ the radius in units of $R_\odot$, and $P$ the pulsation period
in units of day. Therefore, the pulsation mass is given by 

\begin{equation}
M= {{Q^2 R^3}\over P^2}. 
\label{equation_Q}
\end{equation}
Numerical calculations show that 
$Q=0.06$ -- 0.08 for the fundamental mode, and  $Q=0.03$ -- 0.04 for the first 
overtone \citep[e.g.,][]{xio07}. 
Using the radius estimated above and $P=218$ day, 
we can estimate the RG mass (pulsation mass) to be  $M=0.5$ -- $0.9~M_\odot$ for 
both of the fundamental and first overtone modes. 
The $0.8~M_\odot$ is consistent with the visual photospheric radius of 
$335~R_\odot$ in Table \ref{table_RGradius} for the fundamental mode. 

A different way to estimate the RG mass comes  from binary evolution 
theory. A WD of mass  $\sim 0.6~M_\odot$ 
corresponds to a $\sim 2.0~M_\odot$ zero-age main-sequence star in the initial-final 
mass relation derived from observation \citep[e.g., Table 3 in][]{wei00}, or 
a $\sim 3~M_\odot$ in stellar evolution calculation for binaries \citep{ume99}.  
Then, the companion star should be smaller than $ 2-3~M_\odot$,  
because the more massive component in a binary evolves first.  
These values are consistent with the above estimate derived from the pulsation theory.

\subsection{X-ray observation}

PU Vul becomes a supersoft X-ray source in the later phase of the outburst 
when the surface temperature of the WD becomes high enough to emit X-rays.  
\citet{kat11} estimated the supersoft X-ray flux, but it was very 
uncertain because the long-term evolution of the temperature depends on 
the assumed optically-thin mass-loss rate   
as well as the possible absorption due to the RG cool winds. 
In the present work, we confirm that the nuclear burning still continues and 
thus the WD currently evolves toward a supersoft X-ray phase. 
We also showed that the binary system contains two different origins of nebulae, 
i.e., nebula 1 comes from the RG cool-wind and nebula 2 from the   
WD hot-wind.  
This cool-wind origin nebula absorbs a part of the supersoft X-ray flux, because 
the nebula is partially neutral (Rayleigh scattering 
in 1991-1993:\citet{tat09}; Raman scattered \ion{O}{6} lines since 2006: \citet{tat11}). 
Therefore, the supersoft X-ray flux should vary with the binary phase, i.e., the 
flux is minimum when the RG is in front of the WD and maximum when the WD 
is in front of the RG. The $UBV$ light curves of PU Vul \citep{shu11} show   
such a long term variation with the orbital phase.
As mentioned in Section \ref{sec_opticallightcuve} we explain this variation 
as an eclipse of nebula 2 by the RG (and also possibly by the nebula 1). 
Therefore, the supersoft X-ray flux may also show a similar long-term variation. 
We expect that the X-ray flux will be maximum when the $UBV$ flux is maximum.   
If the orbit is circular, the next maximum will be in June 2014, and it 
is a good chance to detect supersoft X-rays.

SMC 3 is a symbiotic star consisting of a massive WD and an M-giant, which 
attracts attention in relation to the progenitor of type Ia supernovae 
\citep{hac10b}. Its supersoft X-ray flux and $B$-magnitude show similar long-term  
variations in the orbital phase \citep{stu11}. 
Sturm et al. analyzed the X-ray variation, but could not give a definite explanation 
about the X-ray variability.  
We could, however, explain this X-ray and $B$-magnitude variations in terms of 
wide eclipses because it is similar to the $UBV$ variations of PU Vul.

\section{Conclusions} \label{sec_conclusions}

Our main results are summarized as follows:

1. We present new estimates of the temperature and radius of the hot component from  
a very early phase of the outburst (1979) until 2011. These are very 
consistent with our theoretical model of outbursting WDs 
based on thermonuclear runaway events without optically thick winds.

2. We analyzed the first (1980) and second (1994) eclipses,  
assuming sinusoidal variations of the brightness and radius of the RG. 
Both of the eclipses are explained as a total eclipse of the WD 
occulted by the pulsating RG. 
Between the first and second eclipses, both of the components shrank 
in size. The radius of the hot component decreases from 
$\sim 100 R_\odot$ to  $\sim 0.1 R_\odot$, which is very consistent with 
our theoretical model. 

3. We are able to construct a composite optical light curve that consists 
of four components of 
emission, i.e., the WD photosphere,  hot nebulae surrounding the WD, 
RG photosphere, and nebulae possibly originating from the RG cool winds. 

4. We have estimated the extinction and distance with various methods, that 
is, the light curve fittings of optical and UV 1590 ~\AA~bands based on 
our WD model, 
direct estimates of color excess, and using $K$-magnitude 
and P-L relation of the pulsating RG companion. 
These different methods yield consistent values of $E(B-V) \sim 0.3-0.4$. 
and $d=4-5$ kpc. We adopt $E(B-V)=0.3$ and $d=0.47$ in the present paper 
as representative values. 

5. We interpret the recent long term evolution of $V$ magnitude 
in terms of eclipse of the hot nebula surrounding the WD: 
the $V$ magnitude gradually decreased from 2002 and 
reached a minimum in 2007 and is now in a recovering phase in 2012. 
This means that hydrogen burning is still ongoing. Therefore,   
we suggest X-ray observations around June 2014 to detect supersoft X-rays.

\acknowledgments
The authors wish to express great thank to S. Shugarov and A. Tatarnikova 
for providing photometric data as well as for discussion on recent 
observations of PU Vul.  
We are very grateful to M. Takeuti and N. Matsunaga for 
providing information and valuable discussion of pulsating red giants. 
We also thank to the anonymous referee for useful comments that helped 
to improve the manuscript. 
We also thank A. Cassatella and R. Gonz\'alez-Riestra
for useful discussion, the American Association of Variable Star Observers (AAVSO) 
and All Sky Automated Survey (ASAS) for archival data of PU Vul.  
This research has been supported in part by the
Grant-in-Aid for Scientific Research (20540227, 22540254) 
of the Japan Society for the Promotion of Science 
and by the Polish Research Grant No. N203 395534.

\end{document}